\PassOptionsToPackage{warn}{textcomp}
\documentclass[sigconf]{acmart}

\def\BibTeX{{\rm B\kern-.05em{\sc i\kern-.025em b}\kern-.08emT\kern-.1667em\lower.7ex\hbox{E}\kern-.125emX}}
    
\copyrightyear{2019}
\acmYear{2019}
\setcopyright{acmlicensed}

\usepackage{times}
\usepackage{graphicx}
\usepackage{verbatim}
\usepackage{url}
\usepackage{color}
\usepackage{alltt}
\usepackage{xspace}

\newcommand{\Comment}[1]{}

\usepackage{epstopdf}
\usepackage{balance}
\usepackage{float}
\usepackage{colortbl}
\usepackage{boxedminipage}

\usepackage{enumitem}

\usepackage{color}

\floatstyle{ruled}
\newfloat{example}{thp}{lop}
\floatname{example}{Example}

\floatstyle{ruled}
\newfloat{readmeexample}{thp}{lop}
\floatname{readmeexample}{ReadmeExample }

\floatstyle{ruled}
\newfloat{logexample}{thp}{lop}
\floatname{logexample}{BuildLogExample }

\floatstyle{ruled}
\newfloat{commandexample}{thp}{lop}
\floatname{commandexample}{CommandExample }

\floatstyle{ruled}
\newfloat{example}{thp}{lop}
\floatname{example}{Example }

\usepackage{listings}
\usepackage[autostyle=false, style=english]{csquotes}
\MakeOuterQuote{"}

\usepackage{amssymb}
\usepackage{graphicx}

\usepackage{multirow}
\usepackage{tcolorbox}

\usepackage{url}
\urldef{\mailsa}\path|{alfred.hofmann, ursula.barth, ingrid.haas, frank.holzwarth,|
\urldef{\mailsb}\path|anna.kramer, leonie.kunz, christine.reiss, nicole.sator,|
\urldef{\mailsc}\path|erika.siebert-cole, peter.strasser, lncs}@springer.com|    

\usepackage{color}

\definecolor{light-green}{rgb}{.5,1,.5}
\definecolor{light-pink}{rgb}{1,0.5,.5}
\usepackage{algorithm,algorithmic}
\usepackage{amsmath}
\usepackage{soul}
\usepackage{subcaption} 
\usepackage{caption} 
\newcommand{\fh}[1]{#1}

\graphicspath{ {figures/} }
\begin{document}

%
\title{An Empirical Study of Software Exceptions in the Field using Search Logs}

%

\author{Foyzul Hassan}
\email{foyzul.hassan@my.utsa.edu}
\affiliation{%
  \institution{University of Texas}
  \streetaddress{}
  \city{San Antonio}
  \state{TX, USA}
}

\author{Chetan Bansal}
\email{chetanb@microsoft.com}
\affiliation{%
  \institution{Microsoft Research}
  \streetaddress{}
  \city{Redmond}
  \state{WA, USA}
}

\author{Nachiappan Nagappan}
\email{nachin@microsoft.com}
\affiliation{%
  \institution{Microsoft Research}
  \streetaddress{}
  \city{Redmond}
  \state{WA, USA}
}

\author{Thomas Zimmermann}
\email{tzimmer@microsoft.com}
\affiliation{%
  \institution{Microsoft Research}
  \streetaddress{}
  \city{Redmond}
  \state{WA, USA}
}

\author{Ahmed Hassan Awadallah}
\email{hassanam@microsoft.com}
\affiliation{%
  \institution{Microsoft Research}
  \streetaddress{}
  \city{Redmond}
  \state{WA, USA}
}


%

%
\begin{abstract}
Software engineers spend a substantial amount of time using Web search to accomplish software engineering tasks. Such search tasks include finding code snippets, API documentation, seeking help with debugging, etc. While debugging a bug or crash, one of the common practices of software engineers is to search for information about the associated error or exception traces on the internet.

In this paper, we analyze query logs from a leading commercial general-purpose search engine (GPSE) such as Google, Yahoo! or Bing to carry out a large scale study of software exceptions. To the best of our knowledge, this is the first large scale study to analyze how Web search is used to find information about exceptions. We analyzed about 1 million exception related search queries from a random sample of 5 billion web search queries. To extract exceptions from unstructured query text, we built a novel and high-performance machine learning model with a F1-score of 0.82. Using the machine learning model, we extracted exceptions from raw queries and performed popularity, effort, success, query characteristic and web domain analysis. We also performed programming language-specific analysis to give a better view of the exception search behavior. These techniques can help improve existing methods, documentation and tools for exception analysis and prediction. Further, similar techniques can be applied for APIs, frameworks, etc.

 
\end{abstract}

%
%

\ccsdesc[500]{Software and its engineering}
\ccsdesc[500]{Information systems~Web search engines}
\ccsdesc[500]{Information systems~Query log analysis}
%
\keywords{software engineering, debugging, web search, machine learning}
%
\maketitle

\section{Introduction}
\label{sec:intro}
With the growing complexity of software systems, use of web search has become ubiquitous in software engineering. More and more software engineers are relying on search engines for various tasks, including finding code snippets, API documentation, debugging, and understanding new concepts ~\cite{stylos2006mica, nasehi2012makes, Sadowski2015}. Prior research on code search~\cite{bajracharya2006sourcerer, Sindhgatta2006, Corley2015} has shown that software engineers depend heavily on search engines for finding information. A study of Google search ~\cite{Sadowski2015} also identified that developers use Web search heavily for code and code-related tasks. Researchers have proposed several approaches ~\cite{Haiduc2013, Martie2017} to improve code search. Despite all these efforts, recent study ~\cite{Rahman:2018:EDU:3196398.3196425} on code-related search behavior identified that code search often requires more effort than more general search intents to find a solution. \fh{Prior study~\cite{Rahman:2018:EDU:3196398.3196425} also identifies that developers tend to search with error and exception message to find solutions, but general-purpose search engines (GPSEs) such as Google, Yahoo! and Bing are better at locating general code issues compared to specific errors or exceptions.} Thus, error or exception search imposes unique challenges for developers to find solutions using GPSEs.

Xin et al.~\cite{Xia:2017} collected Web search logs from 60 developers and interviewed 12 developers to categorize software engineering related search tasks. According to the study, learning new topics is the first most popular task while debugging errors and exceptions is the second most popular task developers accomplish using Web search. Software engineers heavily rely on Web search to not only find documentation for exceptions but also crowd-sourced information from websites like Stack Overflow, GitHub, etc. This underlines the importance of characterizing exception search behavior to be able to improve and build new tools for improving exception debugging using the internet. However, such analysis is non-trivial and yet to be carried out, which motivates our large-scale study of exception search analysis.

In this empirical study, we address two key challenges: 1) Extraction of exceptions from unstructured text, and 2) Leveraging search metadata for characterization of various aspects of exception-related searches (e.g., popularity, effort, success). We collected search logs from Bing.com. a leading GPSE . In GPSE, search queries can be for a wide variety of intents and domains. To overcome the challenge of extracting exceptions from query logs, we propose a novel machine learning model for extraction of exception names (e.g., NoClassDefFoundErrors) and identifiers (e.g., 500) from raw query text. To train and evaluate the machine learning model, we extracted and labeled 348,559 (\textasciitilde0.3 million) search queries \fh{with a semi automatic approach (See Section~\ref{sub:annotation})}. Next, we defined the metrics for characterizing exception search behavior. Prior work on software engineering search analysis ~\cite{Sadowski2015, Rahman:2018:EDU:3196398.3196425,Xia:2017} worked on limited amount of data and data collected in a controlled environment using browser plug-ins or via crowd-sourcing. Such data may not be a good representative of the actual behavior. To overcome this issue, we collected 50 days of user search queries from a leading GPSE and extracted exception names and identifiers with the machine learning model. We also performed Programming Language (PL) specific categorization for better analysis. For behavior analysis, we adopted standard metrics defined and used in the Information Retrieval (IR) ~\cite{castillo2010,Shokouhi2016,Hu2011,akb13-sigir} community and performed behavior analysis based on these metrics. Through this empirical analysis, we address the following research questions:

\begin{itemize}\itemsep6pt
\item \textbf{\emph{RQ1:}} What are the most popular exceptions?\newline
\fh{Seeking help through general-purpose search engines (GPSEs) is a complex process, and to understand developer needs, we need to have a study on how developers interact with GPSEs. Prior study~\cite{Xia:2017} identifies that exceptions are one of the top-most searched items, but the study lacks the detail of exception search items and behavior. Through this research question, we identified mostly searched exceptions that can be more actionable for improving documentation and generate fix suggestions.}
\item \textbf{\emph{RQ2:}} In terms of search effort, what exceptions require the most effort?\newline
\fh{Developers interaction with GPSEs are time intensive and recent study on developer search behavior~\cite{Rahman:2018:EDU:3196398.3196425} identifies that code-related search tends to use more time than non-code related search. In this RQ, we explored search effort for exceptions in terms of time to find an exception solution.}
\item \textbf{\emph{RQ3:}} Which exceptions are most difficult to debug \fh{in terms of search} using Web search?\newline
\fh{While searching for a query, people browse different websites for a satisfactory solution. Through this RQ, we tried to understand the successfulness of various exceptions from different PLs.}
\item \textbf{\emph{RQ4:}} What are the query characteristics of exceptions from various PLs?\newline
\fh{Through this RQ, we tried to answer the search query property:  number of terms in the query. Since for exceptions developers tend to search with exceptions generated by the compiler or framework, it can also help us to identify which exceptions are more verbose in nature.}
\item \textbf{\emph{RQ5:}} Which websites are the most helpful in debugging exceptions?\newline
\fh{In this RQ, we discussed the most frequently used Q\&A sites that are helpful in finding exception solutions. Though developers community considers Stack Overflow, GitHub, etc. as the most used Q\&A sites, this research question can identify other prominent sites for exception solution and the findings can also be helpful for the research community for solution mining and improving the documentation.}
\end{itemize} 


Through answering the above research questions, we characterize the exception search behavior and provide insights for software engineers as well as for researchers. For example, our analysis finds that Python exception search is more effort-intensive than Java and C\# exception search, but have higher search success. Moreover, our analysis on exception search shows that Q\&A sites are more frequently used than official sites for exception solution. Several research studies have been performed to understand how software engineers search for code~\cite{Sadowski2015, Corley2015} and use Web search for various tasks~\cite{Xia:2017, Rahman:2018:EDU:3196398.3196425}. Yet, this is the first empirical study to analyze usage of web search for debugging exceptions. In summary, we make the following contributions:

\begin{itemize}
\item We propose a novel machine learning model that can extract exception names and IDs from unstructured search query text. Based on the evaluation described in Section ~\ref{sub:machinelearningeval}, the model has a high accuracy with a F1 score of 0.82.

\item We analyzed \textasciitilde5 billion web search queries using metrics from the Information Retrieval (IR) community to understand and characterize software exceptions.

\item We conducted the first study to analyze and characterize exceptions by leveraging commercial Web search data. The results provide useful insights for software engineers as well as researchers to provide better tool support and documentation for exceptions.

\item We proposed a novel methodology for analyzing and characterizing exceptions which can also be applied to other software artifacts like APIs, programming frameworks, etc.
\end{itemize}


The rest of the paper is organized as follows: We start by presenting related work and search log terminologies in Section~\ref{sec:related} and Section~\ref{searchlog}, respectively. After that, we discuss the overview of our analysis which includes collecting search query data, extracting exceptions for raw query text, training and evaluation of the machine learning model in Section~\ref{sec:overview}. Section~\ref{sec:empiricalstudy} presents the empirical findings of our study and Section~\ref{sub:application} discussed about possible implications of our proposed extraction model and study. Finally, we discuss threats to validity and conclusion in Section~\ref{sec:threat} and Section~\ref{sec:conclusion}, respectively.

\vspace{-0.3cm}
\section{Related work}
\label{sec:related}

\fh{Web search is heavily used today for various purposes by software engineers such as code search, debugging, downloading, etc. \cite{bansal2019usage}. Prior research works on code search~\cite{bajracharya2006sourcerer,Sim2011, Stolee:2014:SSS:2628068.2581377} identified that developers widely use GPSEs during development. In a survey, Stolee et al.~\cite{Stolee:2014:SSS:2628068.2581377} reported that 85\% of developers perform source code search in GPSEs at least weekly. Similar behavior was also reported by Sim et al.~\cite{Sim2011} that 50\% of developers perform search for code frequently. Moreover, prior study~\cite{Sim2011} identifies that code-specific search engines such as Koder and Krugle~\cite{krugle} perform better searching for subsystems of code, while GPSEs perform better for code blocks.}

\fh{To analyze search behavior, search logs of GPSEs are widely adopted by the different research communities \cite{jhaver2019measuring, bansal2020studying, rao2020product}. Research work~\cite{paparrizos2016detecting} on the healthcare domain utilizes search log for detecting devastating
diseases, while recent work~\cite{chancellor2018measuring} uses search log to measure employment demand. In software engineering research, search logs are also used to analyze search behavior. Bajracharya et al. \cite{Bajracharya2012} analyzed the logs of a code search engine from a 12 months period. They do topic and lexical analysis to understand the usage of code search engines. Further, they compared code search engines with Web search engines and identified aspects unique to code search. Research work to measure developer focus~\cite{Corley2015} used search log of 150 developers over four months period of time. Recent work by Rahman et al.~\cite{Rahman:2018:EDU:3196398.3196425} also utilizes search log to characterize code search behavior over non-code search. To analyze the behavior, they collected search logs from 150 developers that contain nearly 150,000 queries. For the analysis, they collected logs from a controlled environment. In contrast, we collected data from general users to analyze exception search behavior and did the analysis on a very large set of data (5 billion).}

\fh{In a study on Google developers, Sadowski et al.~\cite{Sadowski2015} utilized a combination of survey and log-analysis methodologies to analyze code search behavior. The study identified that programmers frequently search code with an average of five sessions with a total of 12 queries each day. In a recent large-scale study, Xin et al.~\cite{Xia:2017} collected Web search logs from 60 developers and interviewed 12 developers to categorize software engineering related search tasks. They found that exception debugging is the second most frequent task performed by software engineers using web search. In this work, we analyze logs from a Web-scale search engine to do a large scale study of software exceptions. Further, we use an ML based approach to automatically extract exceptions from the raw search query text. To the best of our knowledge, this is the first empirical study to analyze and characterize millions of search queries to extract exceptions and characterize them based on various metrics like popularity, effort and success. Further, similar techniques can be applied for APIs, frameworks, etc.}
\section{Web Search Logs}
\label{searchlog}
For the analysis of exception search behavior, we collected web search logs from a leading commercial general purpose search engine (GPSE). The logs contain a rich set of metadata along with associated click information. The logs are anonymized and do not contain any personally identifiable information (PII) like IP addresses, etc.
\vspace{-0.1cm}
\subsection{Web Search Terminology}
\label{sub:terminologies}
Since we use web search logs for our analysis of exception search behavior, we have adopted some key terms from the web search domain. In this section, we describe the definitions that will be used frequently in the rest of the paper.

\begin{itemize}

\item \textbf{\emph{Search Query:}} A search query is the raw query text entered into the search engine by a user.

\item \textbf{\emph{Search Session:}} For a given client, a search session is defined as a series of search queries that extends until either the browser is closed or there is a 30 minute inactivity~\cite{radlinski2005query,Jiang2015}.

\item \textbf{\emph{Result Urls:}} Ordered list of Urls displayed by the search engine in response to a search query.

\item \textbf{\emph{Clicked Urls:}} List of Urls clicked by the user from the result Urls, ranked based on the order in which they were clicked.

\item \textbf{\emph{Dwell Time:}} Amount of time spent by a user on the clicked result page. Dwell time is considered as one of the key metrics for web search \fh{effort~\cite{castillo2010,Jiang2015,Liu2018}} and has a high correlation with task difficulty and user satisfaction \fh{~\cite{Fox2005,Luo16}}. 

\item \textbf{\emph{SAT (Satisfaction) Click:}} Click Urls with Dwell Time more than 30 sec are considered to be SAT clicks. Prior research on search behavior~\cite{Fox2005, Hassan:2010:BDU:1718487.1718515,Mehrotra2017} adopted SAT Click as signal for relevance of the clicked Url.

\item \textbf{\emph{Search Success:}} Prior work on search analysis ~\cite{Hu2011,Lu18} found that if the last clicked Url answered the user query, they do not explore the search results further. Hence, we consider a search query to be successful, if the last result click for that query is a SAT Click.

\end{itemize}
\section{Methodology}
\label{sec:overview}
\begin{figure*}
  \includegraphics[width=.8\linewidth,height=4.8cm]{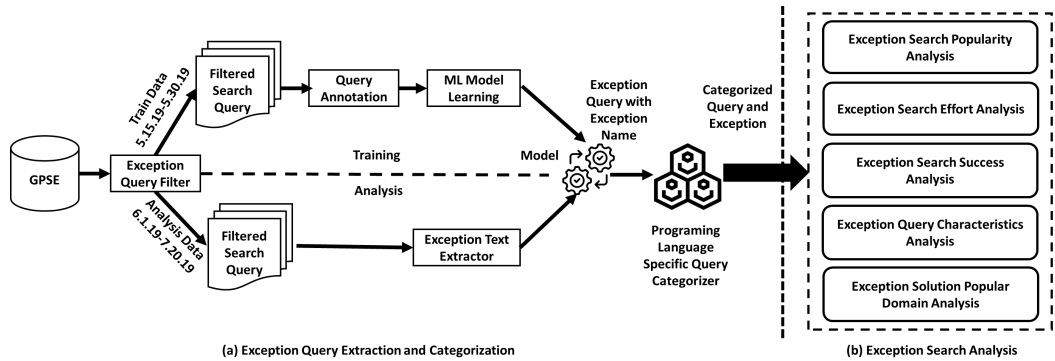}
  \vspace{-0.2cm}
  \caption{Overview of the Study}
  \label{fig:overview}
  \vspace{-0.2cm}
\end{figure*}

In this section, we discussed the two steps of our study: 1) Exception query extraction and categorization, and 2) Exception search analysis. Figure~\ref{fig:overview} shows an overview of the study. Our exception search query extraction process and categorization is described in Section~\ref{sub:classification}. Based on extracted exceptions, we investigate the search behavior to answer RQ1\textasciitilde RQ5 in Section~\ref{sec:empiricalstudy}. We use two datasets from different time periods. The first dataset (May 15-May 30, 2019) is used to train the models to label, extract, and tag the exceptions. The second dataset (June 1-July 20, 2019) is used for the empirical analysis of web search behavior related to exceptions.

\vspace{-0.2cm}
\subsection{Data Filtering}
\label{sub:datafilter}
Web search is ubiquitous in nature and is used in different domains as well by a diverse set of people. Also, the web search patterns varies based on demographics, locales, client, etc. Since in this work we are focusing on exception search behavior, we applied the following filters to remove variation and noise:

\textbf{\emph{Locale \& Region:}} We only used search logs from users with English locale and the US region.  We determined locale-based search request HTTP header~\cite{AcceptLanguage-HTTP}. 
Lastly, We also filtered out search queries that contains non-English characters based on character encoding.

\textbf{\emph{Keywords \& Clicks:}} Since we want to analyze exception related search queries, in order to select error and exception related queries we applied keyword-based filtering. \fh{Prior research works~\cite{SenaMSR16,ZJia2019} on error and exception handling also applied keyword-based search approach to filter related issues or bugs.} For exceptions, based on manual analysis, we used the keywords: error, errno, and exception. \fh{The keyword "error" is used in Python, Javascript, C/C++, Ruby, R, TypeScript as part of exception, while "exception" keyword is used as part of Java, C\#, Php and Perl exceptions. Keyword "errno" is used mainly for ID based exceptions. These keywords are very generic in nature and covers wide range of exceptions from large set PLs.} So, we extracted all the queries which contains at least one of these keywords in either the query text or the clicked Url. Further, to remove noise from the dataset, we filtered out all the search queries which did not result in at least one click. \fh{Prior research works ~\cite{Mao2016, QAi2017,  Azzopardi2020} on search behavior also removed such data to avoid cases where users abandon the search query or remain inactive. This behavior can mislead search behavior analysis.}

\vspace{-0.2cm}
\subsection{Study Subject}
\label{sub:dataset}
We collected search queries from a leading commercial Web search engine such as Google, Yahoo!, Bing, etc.\footnote{Anonymized for double blind} to train and evaluate our machine learning model for extraction of exceptions from raw search query texts. Apart from this dataset, we collected a separate set of search logs from the same Web search engine for exception search behavior analysis.

For \textbf{model training}, we extracted a random sample of search queries from a 15 days period in May 2019. We applied the filters described in Section~\ref{sub:datafilter} to get exception related queries. With the filters applied, we extracted 1158286 (\textasciitilde 1.1 million) search queries along with the associated click information. 

For the \textbf{exception search behavior analysis}, we applied the filters mentioned in Section~\ref{sub:datafilter} to a random sample of 5 billion search queries from 50 days (June 1, 2019 to July 20, 2019) period. Finally, after filtering, we extracted 980155 (0.98 million) search queries from 589050 (0.58 million) distinct search sessions. We mined the session information to perform session-specific analysis.

\vspace{-0.3cm}
\subsection{Exception Query Extraction and Categorization}
\label{sub:classification}
Search engines process search queries from a wide range of domains and intents. Since our research is focused on characterizing exception search behavior, we need to have a mechanism for extracting search queries related to exceptions. In order to be able to extract exceptions from a wide variety of PLs and frameworks, the extraction process should be generic. But due to the large volume and diversity of search queries, hand-crafted regular expressions have low accuracy. Therefore, we developed a high-performance machine learning based approach that can extract exception Ids and names from search queries.

\subsubsection{Query Annotation}
\label{sub:annotation}
For training any machine learning model, one of the main challenge is to have labeled data. Due to wide range exception types from different PLs and frameworks and large volume of search query, annotating exceptions in search query can be effort intensive.  To minimize the effort of training query labeling or annotation, we followed a two step process: 1) Regular expression based labeling, 2) Remove noises from the regular expression based labels. As part of the regular expression based labeling, we hand-crafted regular expressions shown in Table~\ref{table:regex}. We differentiate between ID based exceptions (e.g 404, CS1061) and name based exceptions (TypeError, java.io.IOException). Regular expressions no. 3 mentioned in Table~\ref{table:regex} captures name based exceptions, while other regular expressions capture ID based exceptions.


\begin{table}
\center
\footnotesize
\caption{\label{table:regex} Regular expressions for extracting exceptions from Search Queries}
\vspace{-0.2cm}
\begin{tabular}{|p{0.2cm}|p{3.5cm}|p{3.5cm}|}
\hline
\textbf{No} & \textbf{Regular expression}                                                                                       & \textbf{Sample search query}                                               \\ \hline
1           & (error |errno|err|refused|errorcode|error code|hresult|exit|response|check code|scope|state).*(\textbackslash{}d+) & error 2006 (hy000) at line 462                                                 \\ \hline
2           & (\textbackslash{}d+).*(error|errno|err|refused| errorcode|error code|hresult|exit|response|check code|scope|state) & ssrs 2016 error: an attempt has been...                                         \\ \hline
3           & (?:\textasciicircum{}|{[}, {]}){[}A-Za-z{]}\{1\}{[}A-Za-z.{]}+(error|exception|iteration)
                         & java.lang.TypeNotPresentException: Type javax.xml.bind.JAXBContext not present \\ \hline
4           & 0{[}xX{]}{[}0-9a-fA-F{]}+                                                                                         & 0x800A03EC saveas                                                              \\ \hline
5           & {[}A-Z{]}{[}0-9{]}+;                                                                                              & LNK1189 65535                                                                  \\ \hline
6           & {[}3|4|5{]}{[}0-9{]}{[}0-9{]}                                                                                     & 404 GET /nbextensions/widgets/notebook/js/extension.js                         \\ \hline
\end{tabular}
\vspace{-0.2cm}
\end{table}


Even though we hand-crafted generic regular expressions to extract exceptions from search queries, as expected, it also results in significant amount of false positives. For example, \emph{"Office 2016 error"} is classified as an exception related query and "2016" is as the exception ID. Similarly, \emph{"cyberterror facts"} query also results in a hit with "cyberterror" being tagged as the exception name. To remove these noisy exception queries, we performed manual inspection of exceptions types. For an example, when we applied regular expression based annotation, we found more than thousands "cyberterror" related query as exception query and those are grouped as "cyberterror" exception. With manual analysis we evaluated each exception group and identified that "cyberterror" is not a exception. So, we removed exception annotation of all the queries those are previously identified as exception query by regular expression based approach. In this process, we tried to minimize labeling effort of large number of queries by looking at exception categories, not by each individual search queries.

To create the annotated training dataset for the model, we performed labeling using the regular expressions on the 1.1 million search queries from the training period mentioned at Section~\ref{sub:dataset}. After applying regular expressions based labeling, we collected 348558 (\textasciitilde0.3 million) queries with exception and non-exception query in the ratio 1:1. Two of the co-authors manually validated exception types or groups for training data. We refer to this data as semi-automatically annotated data.

\subsubsection{ML Model for Extracting Exceptions}
\label{sub:machinelearning}
To parse search queries we need to have a machine learning model to extract the exception entity. In natural language processing, an entity is considered as basic information element and often considered the main subject of the text. Named Entity Recognition(NER) is a Natural Language Processing(NLP) technique to identify entities from text and classify them into the defined categories. NER is widely used in different languages processing applications, such as newspaper content classification~\cite{Middleton2016}, Q\&A systems, and machine translation~\cite{hassanawadallah2007improving}, extract software project artifact information from document~\cite{foyzul2017}. NER solutions can be divided into two broad categories: i) rule-based and ii) statistical pattern-based. Rule-based methods are mainly for common entities like persons, locations, organizations, etc. using specialized dictionaries as the reference for identification.  For entities that are not included in the dictionary, may require experts to rewrite the handcrafted rules. On the other hand, the statistical pattern-based approach needs a larger annotated corpus for learning but doesn't need experts. Different supervised machine learning algorithms including HMMs~\cite{collier2000extracting}, SVM~\cite{jonnalagadda2013using}, CRF~\cite{Lafferty2001} have been used for learning of statistical pattern-based NER systems. With our semi-automated tagging approach as described in Section~\ref{sub:annotation}, we annotated the exception entity corpus for NER training.

The conditional random fields (CRF) algorithm proposed by Lafferty~\cite{Lafferty2001} is widely used for reliable sequence labeling ~\cite{Liu2011, Wang09, wei2016disease} for NER task. In our study, we used the first-order Markov linear chain CRF~\cite{Roth2005} with L-BFGS~\cite{lbgs1989} training algorithm using the scaling method~\cite{Rabiner1990}. To regularize the classifier, we used Elastic-net(L1+L2)~\cite{l1l2} penalty in order to reduce model complexity. Based on the Hammersley-Clifford theorem, CRF computes the conditional probability of a state sequence as follows:
\begin{equation}
p_\theta(\mathbf{y}|\mathbf{x}) = \frac{1}{Z_\theta(\mathbf{x})} \exp \Bigg\{\sum_{t=1}^T \sum_{k=1}^K \theta_k  f_k(y_{t-1}, y_t, x_t) 
\Bigg\}
\label{eq:crf}
\end{equation}
\vspace{0.2cm}
where $\mathbf{x} = (x_1,\dots, x_T)$ denotes the input sequence and $\mathbf{y} = (y_1,\dots, y_T)$ is the output sequence, hereafter referred to as the sequence of labels. $\{f_k\}_{1\leq k \leq K}$ is an arbitrary set of feature functions and 
$\{\theta_k\}_{1\leq k \leq K}$ are the associated real-valued parameter values. For our study, we used CRFSuite~\cite{CRFsuite}, which is a commonly used Python library for CRF. As part of feature extraction, tokens and parts of speech (POS) tags are the basic features to extract. The basic features have less contextual information and less text patterns. So, in addition to basic features, we extract three other features, including contextual features ~\cite{fayolle2010crf}, gazetteer features ~\cite{carlson2009learning}, and orthographic features ~\cite{ma2016end}. For hyper-parameters, we used 0.1 as Elastic-net L1 penalty and 0.01 as Elastic-net L2 penalty based on prior work on domain-specific parser~\cite{parserator}. Due to large data size, we used maximum iteration count 200 for training.

\subsubsection{Model Evaluation}
\label{sub:machinelearningeval}
In this section, we describe the evaluation of the performance of CRF model described in Section~\ref{sub:machinelearning}. We evaluateed the model using a \emph{manually annotated dataset}. For this dataset, we randomly sampled 500 queries with exception and non-exception queries in the ratio 1:1 collected from the analysis period (June 1, 2019 to June 5, 2019). Two researchers separately annotated those queries and resolved the disagreement by discussion. We performed Cohen's kappa coefficient~\cite{kappa} to find inter-annotator agreement. Coefficient value 1 indicates a complete agreement and value of 0 indicates complete disagreement. In our annotation, we found the coefficient value as 0.88, which indicates high confidence of agreement. 

\begin{table}[]
\footnotesize
\caption{\label{table:evaluation} CRF model performance evaluation.}
\vspace{-0.4cm}
\begin{tabular}{|l|l|l|l|l|}
\hline
\textbf{Class} & \textbf{Precision} & \textbf{Recall} & \textbf{F1-Score} & \textbf{Support} \\ \hline
Exception ID   & 0.89               & 0.69            & 0.78              & 169              \\ \hline
Exception Name & 0.78               & 0.97            & 0.86              & 61               \\ \hline
Avg.           & 0.83               & 0.83            & 0.82              &                  \\ \hline
\end{tabular}
\vspace{-0.4cm}
\end{table}

Since we categorized exceptions into two broad categories: i) ID-based Exception and ii) Name-based exception, we evaluated the performance of both types of category. Table~\ref{table:evaluation} shows the performance of the machine learning model on manually labeled data. According to the Table, average Precision, Recall, and F1-Score are 0.83, 0.83 and 0.82 respectively. Since our evaluation performed with a dataset collected from real-world user search data and large scale in nature, the performance of our model is reasonable and can be used for exception data collection for empirical analysis of exception search behavior.

\subsubsection{Exception Tagging for Analysis}
\label{sub:mlexceptiondata}
Using the machine learning model described in Section ~\ref{sub:machinelearning}, we tag the analysis dataset. If the model can extract any exception from the query then we consider that as an exception query. In many cases, search queries can have multiple exceptions IDs or names due to search exception trace from IDEs. In those cases, we only consider the root or first level of exception. With the machine learning model, we extracted 118315 ID-based exception query and 32887 Name-based exception query from 980155 search queries.

\subsubsection{Programming Language Categorization}
\label{sub:progcategorization}
Every programming language (PL) can have its own format for exception names and Ids. To perform PL specific exception analysis, we categorized search queries into various PLs based on the exception, query text and the clicked Urls. For this, we picked three popular programming languages 1) Java, 2) C\#, and 3) Python. To categorize search queries into these programming languages, we performed a keyword-based search with keywords (java, c\#, and python). We search the keywords in raw query and also in the clicked Urls. If match is found then we assign the corresponding PL to the query. \fh{For instance, for "Search query: python ImportError: cannot import name" we found the PL name was mentioned in the query. So, we assigned this exception query to Python PL category. If PL name was not found in the search query or the clicked Urls; we look-up the exception tagged in the search query in exception lists for Java ~\cite{javaexceptions}, C\# ~\cite{csharpceptions} and Python ~\cite{pythonexceptions}. If we find a match with any of these lists, then we assign the corresponding PL name to that query. For example, in query "Search query: System.io.filenotfoundexception addinutil.exe " , no PL name occurs in the search query or the clicked Urls.} So we cross-reference exception name with the C\# exception list and found a match. So, we assigned C\# as PL type for that query. If we do not find keyword-based match or exception list based matches, we don't assign any PL name for the query. With this approach, we categorized 17035 exception search queries as Java related, 13452 exception search queries as  C\# related, and 27723 queries as Python related.


\vspace{-0.2cm}
\subsection{Exception Analysis}
\label{sub:progcategorization}
Based on the exceptions and PL tagging of search queries, we performed empirical analysis on exception search behavior. We analyzed the following five aspects: 1) Popularity Analysis, 2) Effort Analysis, 3) Success Analysis, 4) Exception Query Characteristics and 5) Website analysis for debugging of exceptions. The results of the study are presented in Section~\ref{sec:empiricalstudy}.
\vspace{-0.2cm}
\section{Empirical Study}
\label{sec:empiricalstudy}
In this section, we present our analysis of exception search queries. Some exceptions can be rare or their search results could not have yielded a meaningful result. To avoid such unusual cases, we only considered exceptions that appeared in at least twenty distinct search sessions. 

\subsection{RQ1: What are the most frequently searched exceptions?}
\label{subsec:rq1}

\subsubsection{Metric}
To find frequently searched exceptions, we used unique session count as a metric as the same exception can be searched for with different text.

\subsubsection{Analysis}
\begin{figure} 
    \vspace{-0.2cm}  
    \includegraphics[width=.8\linewidth,height=4.5cm]{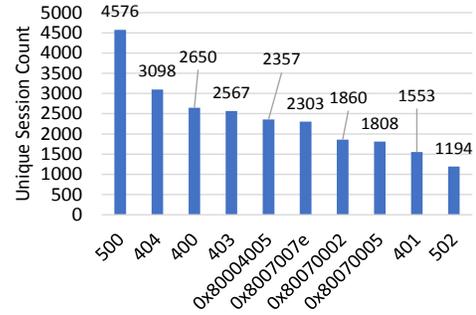}
    \vspace{-0.4cm}
    \caption{Popular ID exceptions}    
    \label{popularidexp} 
\end{figure}

\begin{figure}
    \vspace{-0.2cm}
    \includegraphics[width=.8\linewidth,height=4.7cm]{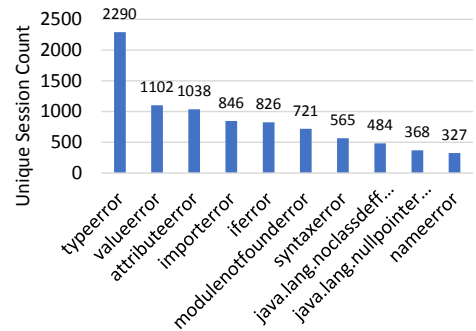}
    \vspace{-0.4cm}
    \caption{Popular name exceptions}
    \vspace{-0.3cm}
    \label{popularnameexp}
\end{figure}

Exceptions can be presented in two text formats: ID-based exception and Name-based exception. We measure the frequency of both ID-based exceptions and Name-based exceptions. Based on unique session count, Figure~\ref{popularidexp} shows the most frequently searched ID-based exceptions. Among the top ten ID-based exception list, six of the exceptions (500, 404, 400, 403, 401, and 502) are Http protocol exceptions and rest are Windows OS related exceptions.  Figure~\ref{popularnameexp} shows most frequently searched Name-based exceptions. In Name-based exceptions, typeerror is the most searched exception. A reason might be due to the fact that  both Python and JavaScript throw \emph{typeerror} exception. Other popular Name-based exceptions are also related to Python and Java.

\begin{figure*}[h]
  \begin{subfigure}[b]{0.32\textwidth}
    \includegraphics[width=\textwidth,height=4.8cm]{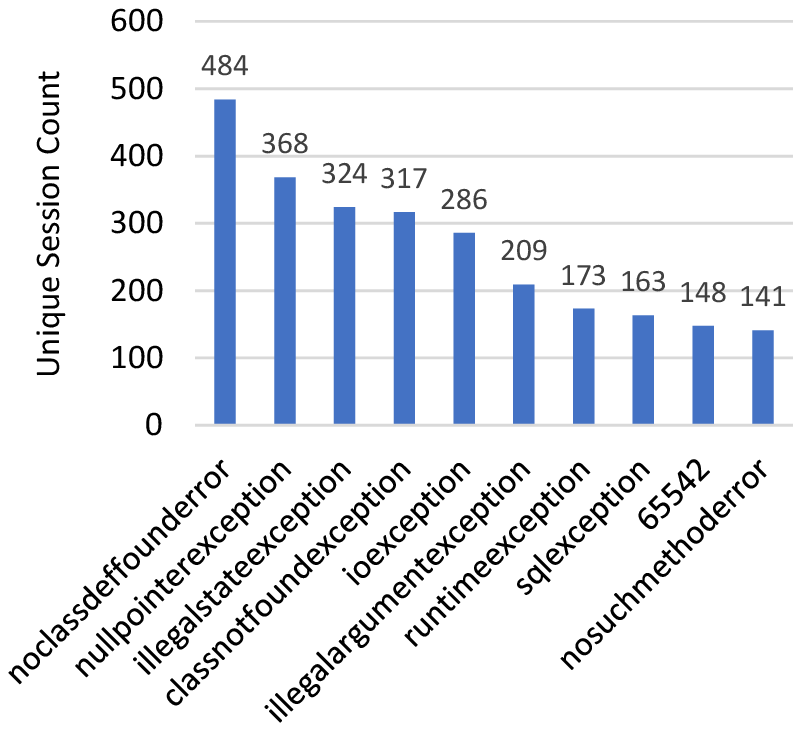}
    \caption{Java}
    \label{popularjava}
  \end{subfigure}
  \begin{subfigure}[b]{0.32\textwidth}
    \includegraphics[width=\textwidth,height=4.8cm]{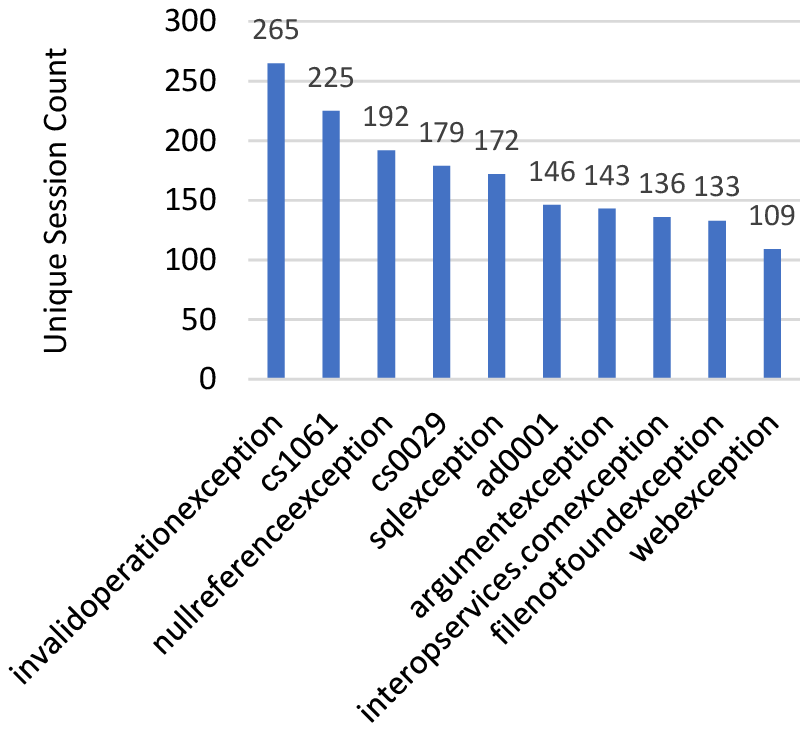}
    \caption{C\#}
    \label{popularcsharp}
  \end{subfigure}
    \begin{subfigure}[b]{0.32\textwidth}
    \includegraphics[width=\textwidth,height=4.8cm]{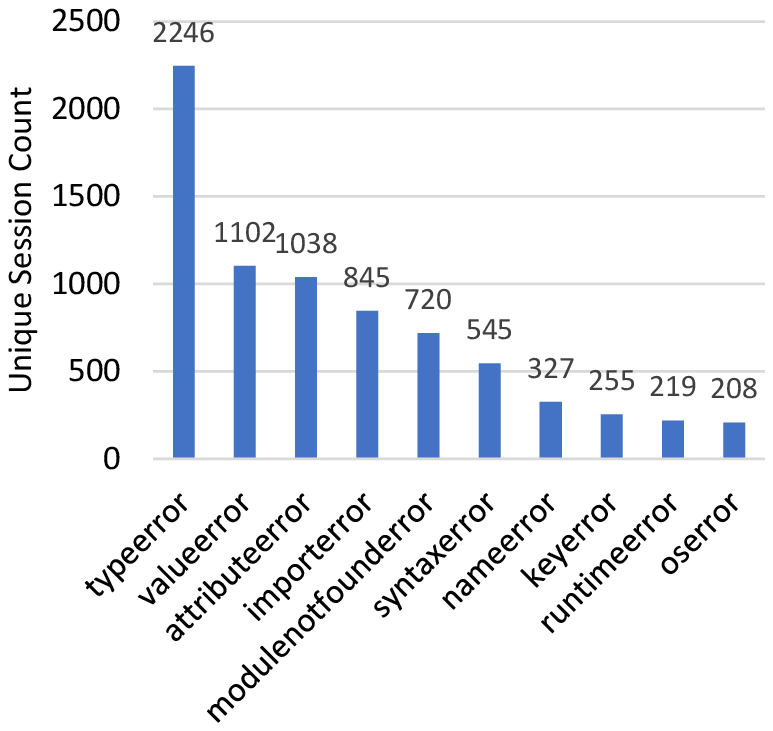}
    \caption{Python}
    \label{popularpython}
  \end{subfigure}
  \vspace{-0.3cm}
  \caption{PL specific most popular exceptions}
  \vspace{-0.3cm}
  \label{popularpl}
\end{figure*}

We also performed programming language (PL) specific popularity analysis. Figure~\ref{popularpl} shows the top searched exceptions of Java, C\#, and Python. For Java, \emph{noclassdeffounderror} is the most popular exception. Another exception we would like to call attention to is exception \emph{65542} which is thrown from the Java utility library for use with OpenGL. For C\#, \emph{invalidoperationexception} is the most frequently searched exception. Even though \emph{cs1061} and \emph{cs0029} both are C\# compile time exceptions, they are also frequently searched. Also, the C\# exception \emph{ad0001} that is thrown from code analyzer also shows up in the top searched for exceptions. For Python, \emph{typeerror} is the mostly searched exception, which is raised when an operation or function is applied to an object of inappropriate type. Others frequent exceptions are also from Python built-in exceptions. Python's \emph{typeerror} exception is searched 4.64 times more frequently than Java's most frequent exception and 8.47 times more frequently than C\#'s top searched exception. This also indicates that Python exceptions are more frequently searched than Java and C\#. Prior studies ~\cite{Nanz2015,Durieux2019} on programming language popularity found that Python is more popular than Java and C\#, which aligns with our exception search popularity findings.



\begin{tcolorbox}
	\textbf{Finding 1:}
	\emph{Most popular ID-based exceptions are related to HTTP Connection and OS related errors. Python exceptions are more frequently searched than Java and C\# exceptions.}
\end{tcolorbox}
\vspace{-0.4cm}

\subsection{RQ2: What is the exception search behavior based on effort?}
\label{subsec:rq2}
\subsubsection{Metric}
For effort analysis, we use Total Dwell Time (described at Section~\ref{sub:terminologies}) in seconds as the evaluation metric. We limit Dwell Time in a Url to 600 seconds to avoid the case where the user clicked the Url, but remained inactive over 600 seconds.

\subsubsection{Analysis}
\begin{figure}[H]
  \includegraphics[width=.8\linewidth,height=4.8cm]{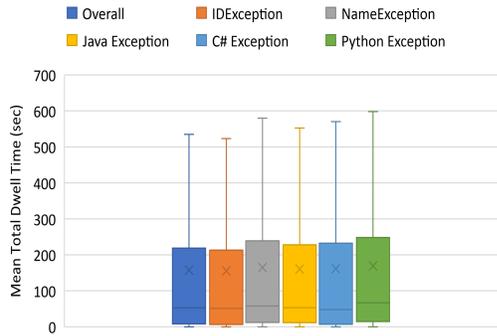}
  \caption{Effort comparison}
  \label{fig:effortall}
\end{figure}

Our exception search effort analysis is divided both by exception format (ID/Name) as well as programming language. According to Figure~\ref{fig:effortall}, overall exception search takes an average effort of 157.39 sec. While for ID-based exceptions it takes 155.22 secs, Name-based exceptions take 164.57 secs of effort, which is 6.02\% higher than ID-based exception search. Name-based exceptions are more effort-intensive than ID-based exception (confirmed by t-test with p-value(ID,Name) = $8.86e-13$, which is lower than threshold 0.05). This could be due to the fact that IDs are usually unique are more easily "searchable" due to easy matching. For programming language specific effort analysis, Java and C\# take a mean effort time of 160.59 sec and 161.57 sec respectively. While Python exception searches take 169.18 sec, which is 5.34\% higher than Java and 4.71\% than C\#. Based on our analysis Python exceptions are most effort-intensive exceptions(confirmed by t-test with p-value(Java,Python) = $0.002$ and p-value(C\#,Python) = $0.03$, which are lower than threshold 0.05). Java and C\# exceptions take similar effort for search which we confirmed by t-test with p-value(Java,C\#) = $0.79$. Prior work~\cite{Fox2005} on effort with Dwell Time finds that it has high correlation with task difficulty and user satisfaction.

\begin{figure*}[h]
  \begin{subfigure}[b]{0.32\textwidth}
    \includegraphics[width=\textwidth,height=4.8cm]{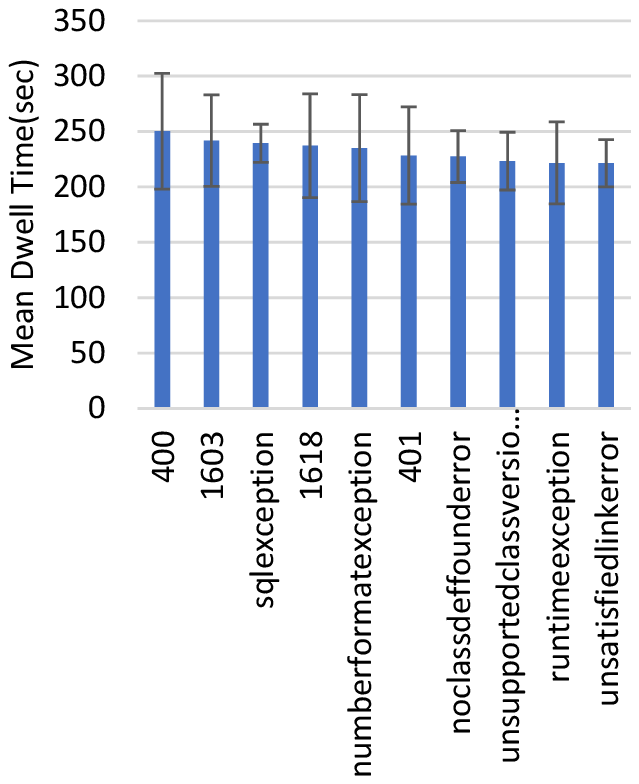}
    \caption{Java}
    \label{effortjava}
  \end{subfigure}
  \begin{subfigure}[b]{0.32\textwidth}
    \includegraphics[width=\textwidth,height=4.8cm]{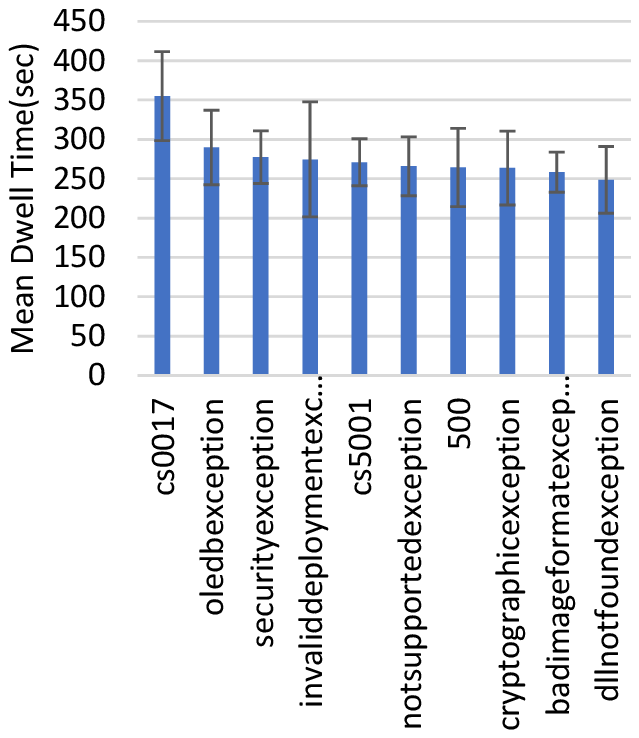}
    \caption{C\#}
    \label{effortcsharp}
  \end{subfigure}
    \begin{subfigure}[b]{0.32\textwidth}
    \includegraphics[width=\textwidth,height=4.8cm]{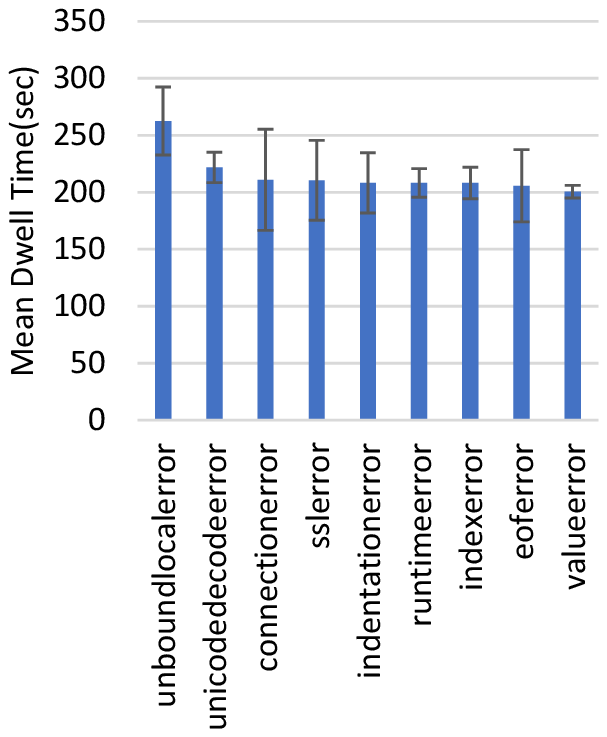}
    \caption{Python}
    \label{effortpython}
  \end{subfigure}
  \caption{PL specific exceptions requiring most effort}
  \label{effortpl}
\end{figure*}

Figure~\ref{effortpl} shows most effort-intensive exceptions of Java, C\#, and Python. Even though \emph{400} is not a Java built-in exception, Java code may generate \emph{400} exceptions for Http protocol usage from code. Other interesting Java exceptions are \emph{1603} and \emph{1618}, which are Java run-time and deployment exceptions. In terms of C\#, cs0017 is the most search effort-intensive exception and the surprising factor for this exception is that it is a compile-time exception due to more than one application entry point. Defining one entry point for an application is a basic concept of any programming language. But our analysis identifies that this the most intensive exception search for C\# and sheds light that better documentation for this could help alleviate problems for developers. For Python, \emph{unboundlocalerror} is the most effort-intensive exception search which is raised when a reference is made to a local variable in a function or method, but no value has been bound to that variable.

For the top effort-intensive Java exceptions, search effort is at least 37.83\% higher than the average search effort for Java exceptions. For C\#, top effort-intensive exceptions, this value is 58.17\% higher than average C\# exception search effort. Top Python search effort-intensive exceptions are 15.49\% higher than average Python exception search effort. The high variance in effort for Java and C\#'s top effort-intensive exceptions indicates that there could be dissatisfaction with the search results or difficulty in these exceptions.
\begin{tcolorbox}
	\textbf{Finding 2:}
	\emph{ID-based exceptions take less effort to find a solution. In terms of PL specific analysis, C\# exceptions and Java takes similar effort, while Python takes higher effort than Java and C\#.}
\end{tcolorbox}

\subsection{RQ3: What is the exception search behavior based on success?}
\label{subsec:rq3}
\subsubsection{Metric}
Search success can be defined as user found information for a query that the user entered. To evaluate exception search success behavior, we will use the Search Success metric which we defined in Section~\ref{sub:terminologies}. The value for this metric is either 0 for fail and 1 for success.

\subsubsection{Analysis}
\begin{figure}[H]
  \vspace{-0.4cm}
  \includegraphics[width=.8\linewidth,height=4.5cm]{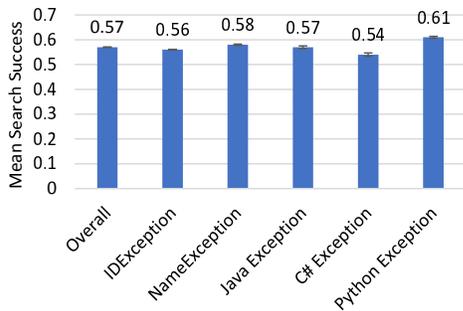}
  \vspace{-0.4cm}  
  \caption{Exception search success comparison}  
  \label{fig:successall}
  \vspace{-0.3cm}  
\end{figure}

For exception search success analysis, we analyzed the overall dataset of exceptions, format-based exceptions and PL specific exceptions. Figure~\ref{fig:successall} shows the mean success rate of all the groups of exceptions. The overall success rate for exceptions is 0.57, while ID-based exception shows the mean success rate of 0.56 and Name-based exception shows the mean success rate of 0.58. So, ID-based exceptions are less successful than Name-based exception in terms of finding a solution from search engines(confirmed by t-test with p-value(ID, Name) = $2.45E-13$, which is lower than threshold 0.05). In terms of PL specific analysis, Python exceptions are the most successful in terms of finding a solution from the search engines with mean success rate 0.61. Among the three programming languages, C\# shows least success rate which 0.54. Mean success rate of these three programming languages are also statistically significant with t-test p-Value(Java,C\#) = $2.05e-31$, p-Value(Java,Python) = $0.01$, and p-Value(C\#,Python) = $1.35e-98$.

\begin{figure*}[h]
  \begin{subfigure}[b]{0.32\textwidth}
    \includegraphics[width=\textwidth,height=4.8cm]{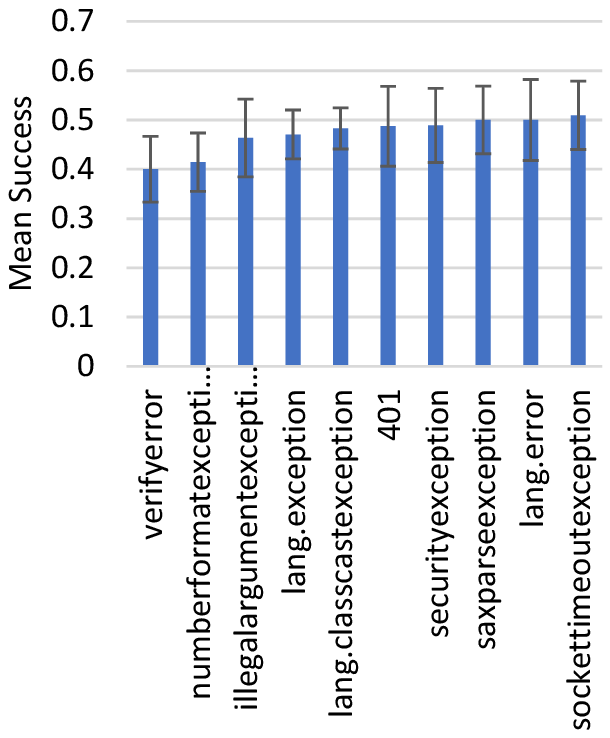}
    \caption{Java}
    \label{successjava}
  \end{subfigure}
  \begin{subfigure}[b]{0.32\textwidth}
    \includegraphics[width=\textwidth,,height=4.8cm]{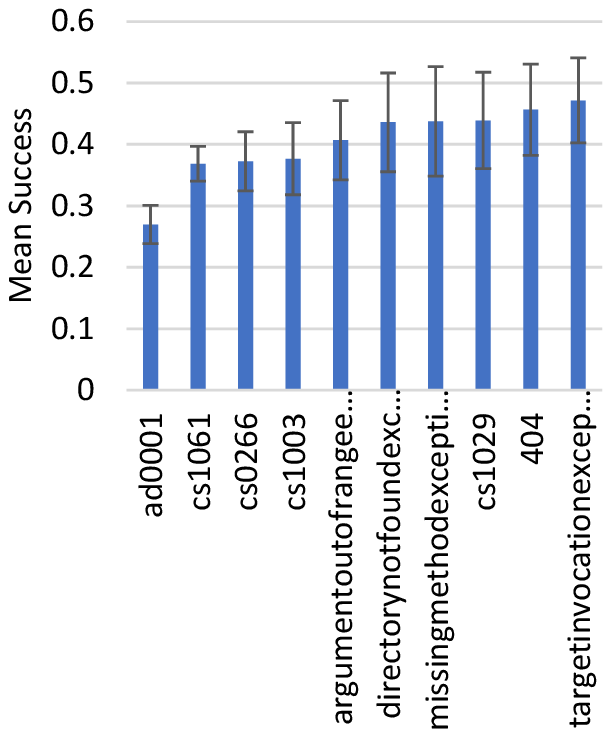}
    \caption{C\#}
    \label{successcsharp}
  \end{subfigure}
    \begin{subfigure}[b]{0.32\textwidth}
    \includegraphics[width=\textwidth,,height=4.8cm]{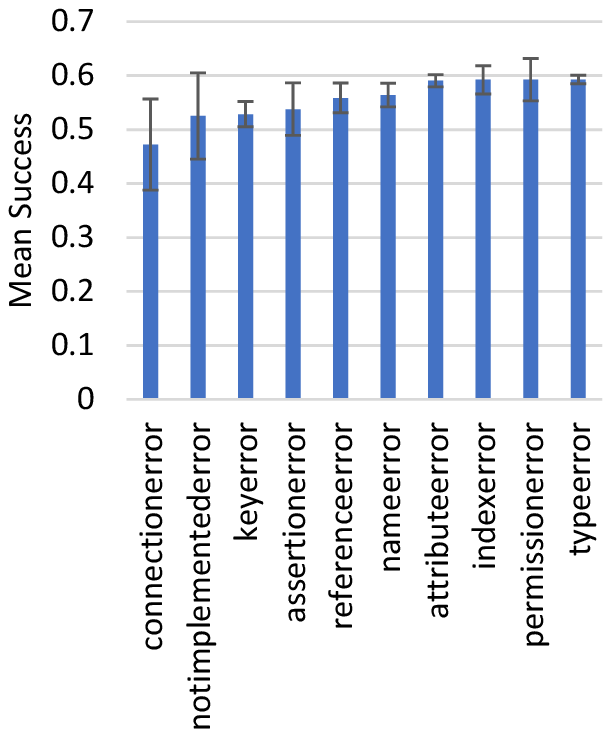}
    \caption{Python}
    \label{successpython}
  \end{subfigure}
  \caption{PL specific exceptions with least success}
  \label{successpl}
\end{figure*}

Apart from group-specific success analysis, we also performed analysis to find the least successful exceptions of Java, C\#, and Python. According to Figure~\ref{successpl}, \emph{verifyerror} is the least successful exception among Java exceptions. \emph{java.lang.VerifyError} can occur when the compile-time and run-time environments are different. Another less successful exception of Java is \emph{saxparseexception}, which might be caused during XML parsing and the solution also depends on XML content. For C\#, top four least successful exceptions are due to code analysis (\emph{ad0001}) and compile-time exceptions (\emph{cs1061}, \emph{cs0266}, and \emph{cs1003}). This might be an indication that the C\# development environment finds a lot more exceptions during code analysis and compile-time than before the run-time executions, which is good from a deployment perspective. For Python, \emph{connectionerror} is the least successful exception and it is also an exception that depends on the connection environment rather than on the code alone.

\begin{tcolorbox}
	\textbf{Finding 3:}
	\emph{ID based exceptions are less successful than Name-based exceptions. Also, among the three Programming Languages C\# has the lowest search success. }
\end{tcolorbox}

\vspace{-0.3cm}
\subsection{RQ4: Characterizing exception search queries}
\label{subsec:rq4}
\subsubsection{Metric}
We used the number of words or terms in the exception search query as the query characteristics. To count number words we tokenize the raw query into words with space used as the delimiter.

\subsubsection{Analysis}
\begin{figure}
  \vspace{-0.2cm}
  \includegraphics[width=0.8\linewidth,height=4.6cm]{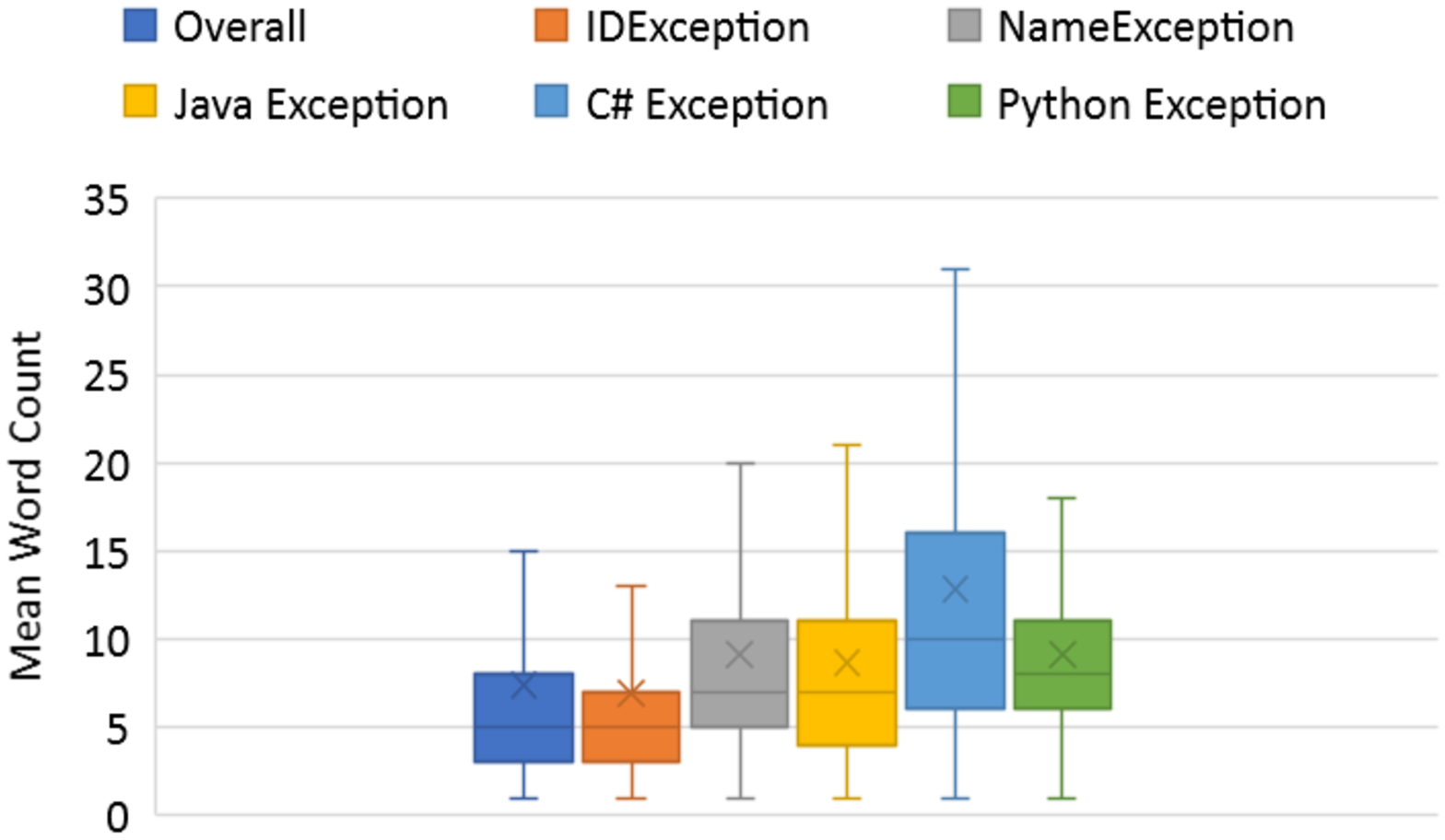}
   \vspace{-0.4cm}
   \caption{Word count comparison for exception search queries} 
  \label{fig:wordcountall}
  \vspace{-0.4cm}
\end{figure}

Figure~\ref{fig:wordcountall} shows the word count characteristics of overall all exceptions, \emph{IDException}, \emph{NameException}, and the three PL specific exceptions. As shown in the graph, ID-based Exceptions are least wordy due to the uniqueness of the exception code. ID-based exception mean word count is 6.93, while Name-based exception word count 9.13 (confirmed by t-test with p-value (ID,Name) = $0.0$, which is lower than threshold 0.05).  C\# exceptions mean word count is 12.72, Java mean word count is 8.53, and Python mean word count is 9.25. Among the three programming languages, C\# exceptions are the most verbose and Java is the least verbose, which is also confirmed by t-test with p-Value (Java,C\#) = $2.37e-141$, $p-Value (Java,Python) = 3.93e-07$, p-Value(C\#,Python) = $1.95e-213$. Users search exceptions with exception text message generated by the compiler or during run-time. This indicates that C\# compiler or run-time generates more wordy exception message than Java and Python.

\begin{figure*}[h]
  \begin{subfigure}[b]{0.32\textwidth}
    \includegraphics[width=\textwidth,height=4.7cm]{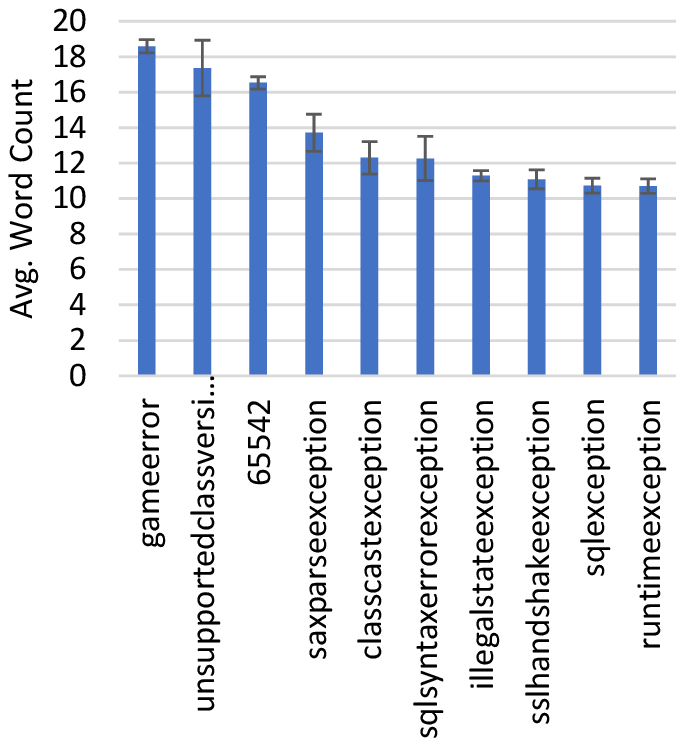}
    \caption{Java}
    \label{fig:wordjava}
  \end{subfigure}
  \begin{subfigure}[b]{0.32\textwidth}
    \includegraphics[width=\textwidth,height=4.7cm]{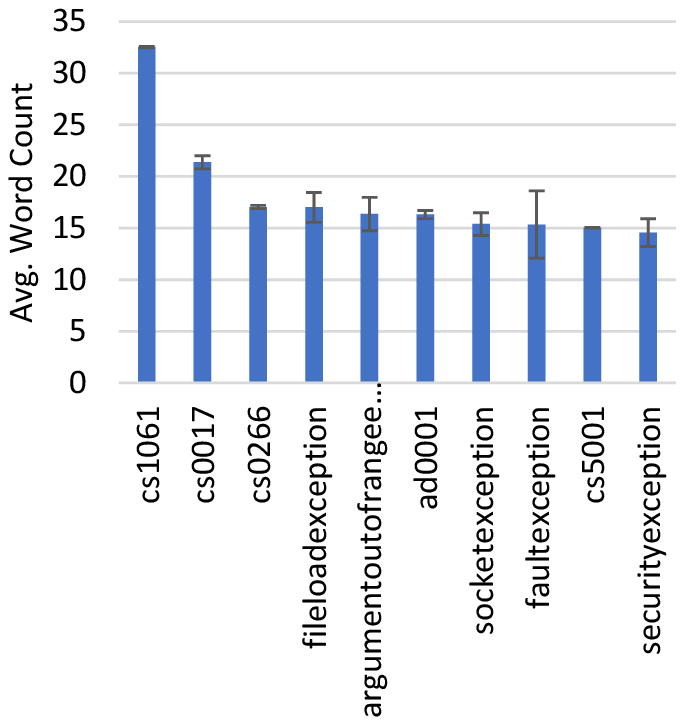}
    \caption{C\#}
    \label{fig:wordcsharp}
  \end{subfigure}
    \begin{subfigure}[b]{0.32\textwidth}
    \includegraphics[width=\textwidth,height=4.7cm]{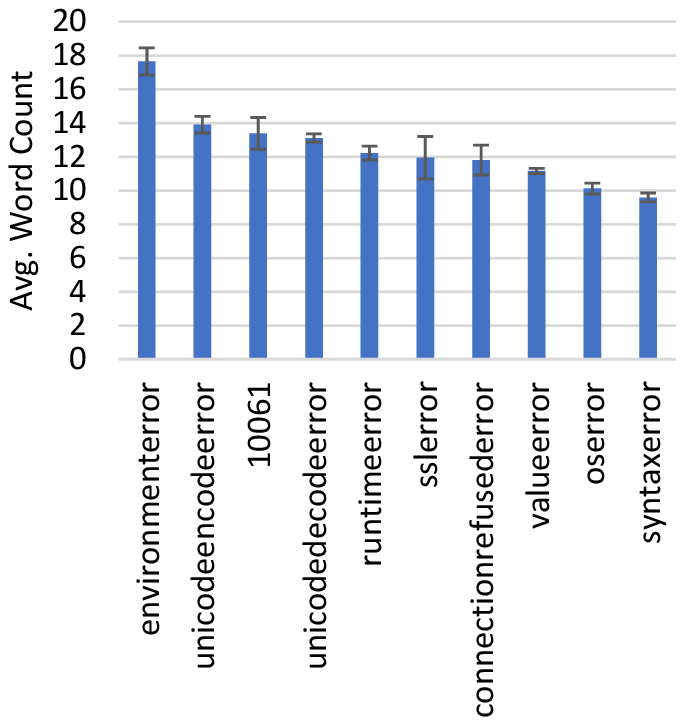}
    \caption{Python}
    \label{fig:wordpython}
  \end{subfigure}
   \vspace{-0.4cm}
  \caption{Most verbose exceptions from Java, C\# and Python}
  \vspace{-0.2cm}
  \label{plwordcount}
\end{figure*}

Figure~\ref{plwordcount} shows the most verbose exceptions of Java, C\#, and Python. Even though \emph{gameerror} is not Java's built-in exception, it's the wordiest exception of Java. For C\#, \emph{cs1061} is the most verbose exception that is thrown when trying to call a method or access a class member that does not exist. For the case of Python, \emph{environmenterror} which is the base class of \emph{IOError}, \emph{OSError} exception is the wordiest exception.

\begin{tcolorbox}
	\textbf{Finding 4:}
	\emph{C\# generates more verbose exceptions than Java and Python.}
\end{tcolorbox}

\subsection{RQ5: What are the popular web domains for finding solutions to exceptions?}
\label{subsec:rq5}
\subsubsection{Metric}
For this analysis, we used the metric click count to a certain web domain for analysis and ranking of its helpfulness in finding solutions to exceptions.

\subsubsection{Analysis}
\begin{figure}[H]
  \vspace{-0.4cm}
  \includegraphics[width=0.8\linewidth,height=4.8cm]{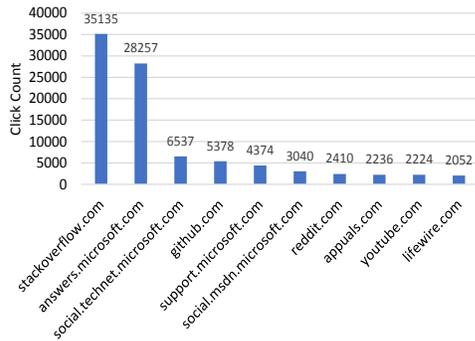}  
  \caption{Most popular websites used for debugging exceptions}  
  \label{fig:populardomain}
  \vspace{-0.5cm}
\end{figure}

Figure~\ref{fig:populardomain} shows the most popular exception solution web domains. According to our analysis, stackoverflow.com is the topmost web domain to provide a solution or help with exceptions. The next popular web domain is from Microsoft community help site answers.microsoft.com site which covers troubleshooting help for a wide range of Microsoft products. Github is the fourth most popular web domain for helping find solutions to exceptions.

For the Java programming language, \emph{stackoverflow.com} is the most popular web domain which is 17.44 times more popular than the Java official community site. \emph{stackoverflow.com} is also the most popular solution web domain for C\#, even though Microsoft maintains several forums and official help sites. For Python exceptions, \emph{stackoverflow.com}  is 35.79 times more popular than the official Python forum. The analysis on Web domain help sites indicates that even though each PL maintains official documentation and community sites, \emph{statoverflow} due to its community effort dominates against them.

\begin{figure*}[h]
  \begin{subfigure}[b]{0.32\textwidth}
    \includegraphics[width=\textwidth,height=4.8cm]{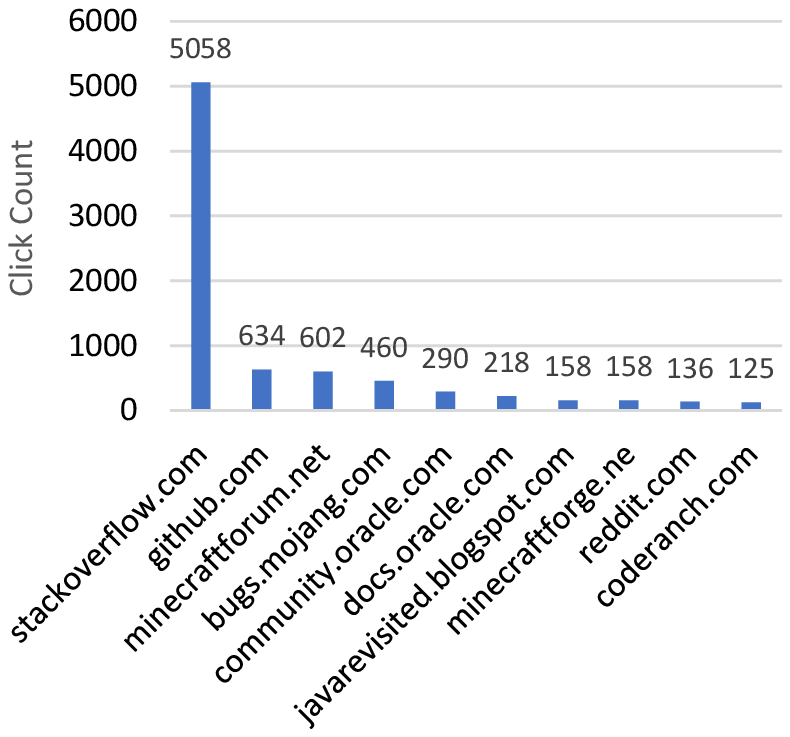}
    \caption{Java}
    \label{popularjavadomain}
  \end{subfigure}
  \begin{subfigure}[b]{0.32\textwidth}
    \includegraphics[width=\textwidth,height=4.8cm]{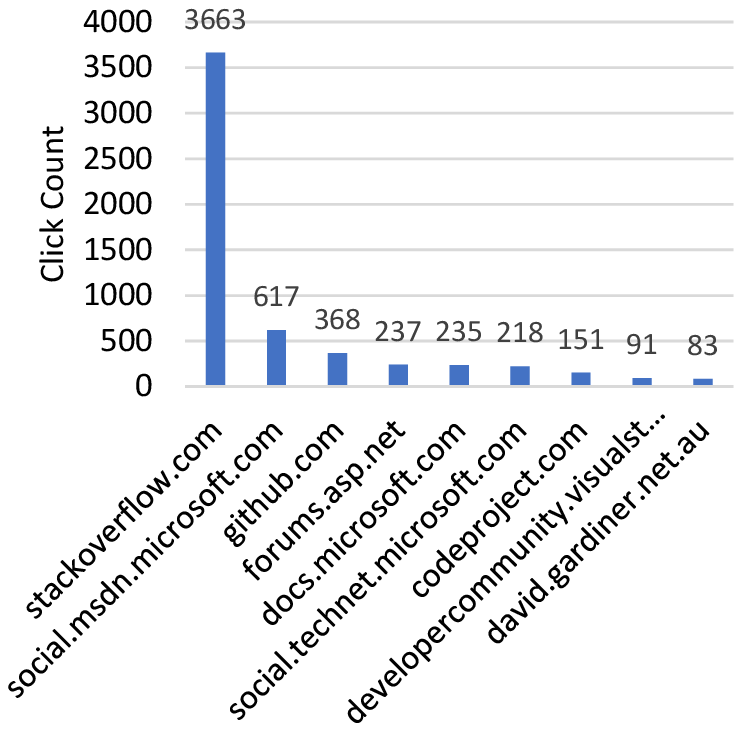}
    \caption{C\#}
    \label{popularcsharpdomain}
  \end{subfigure}
    \begin{subfigure}[b]{0.32\textwidth}
    \includegraphics[width=\textwidth,height=4.8cm]{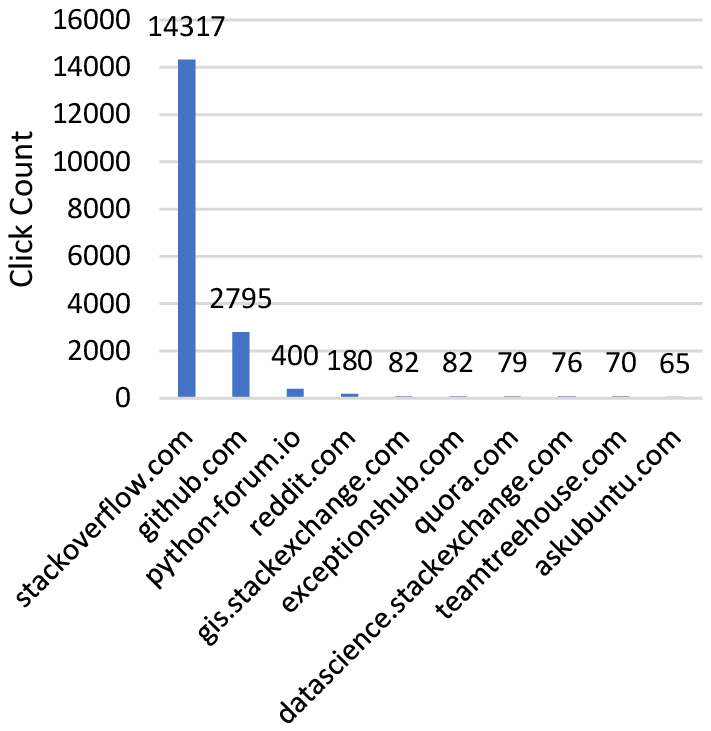}
    \caption{Python}
    \label{popularpythondomain}
  \end{subfigure}
  \vspace{-0.4cm}
  \caption{PL specific popular websites for exception debugging}
  \vspace{-0.25cm}
  \label{popularpldomain}
\end{figure*}

\begin{tcolorbox}
	\textbf{Finding 5:}
	\emph{Even though Java, C\#, and Python have their own documentation websites and social forums, stackoverflow.com is the most popular website for debugging exceptions in these languages.}
\end{tcolorbox}
\vspace{-0.2cm}
\section{Implications}
\label{sub:application}
In this section, we discuss the implications of our exception extraction and categorization model and exception search behavior analysis. Followings are the list of actionable items inferred from the study:

\begin{itemize}
\item \fh{In recent works~\cite{Zhang2016,Chen2017,Kim2018}, researchers have been working on code recommendation systems to assist developers in writing code faster. These works mainly focus on APIs. Through RQ1, we identified the most frequently searched items from different PLs. Also, we identified that ID-based exceptions such as 400, 500, etc. are searched frequently.  These frequently searched items can be base for recommending code for exception solution. From our analysis, we also found several Windows errors, such as 0x80004005, searched very frequently. These type of analysis can be helpful to identify emerging issues such as OS upgrade problem, framework version synchronization problem, etc.}

\item \fh{Analysis from RQ2 indicates that ID-based exceptions take less effort or time to find a solution than Name-based exception. In terms of PL specific analysis, C\# exceptions and Java takes similar effort, while Python takes higher effort than Java and C\#. RQ3 indicates that ID based exceptions are less successful than Name-based exceptions. Also, among the three Programming Languages, C\# has the lowest search success. Prior work~\cite{Fox2005} on search effort analysis finds that effort has a high correlation with user satisfaction. That indicates that C\# exceptions searches are less successful, so developer tends to put less effort or time for searching and abandon search at some point. This is a very critical finding to have the necessity to improve C\# exception handling documentation and posts in Q\&A sites. The same issue is also identified for ID-based exceptions.}

\item \fh{Analysis from RQ4 indicates that C\# generates more verbose exceptions than Java and Python. According to the recent study on web search~\cite{Xia:2017}, developers search though exception messages generated by the compiler or framework with partial message from client code. Prior study~\cite{Rahman:2018:EDU:3196398.3196425} finds that code-related searches are more verbose than non-code related searches. So, the exception search success rate for C\# should be higher than the other two programming languages. But RQ3 analysis finds that the success rate of C\# search is lower than Java and Python. Findings of these two RQs sheds light on the necessity of reviewing the context and information generated by C\# framework exceptions.}

\item \fh{Although several research works~\cite{Kim2018,B_Vasilescu2013,Nguyen2018,Meng2018,Rastogi2018} have used \emph{stackoverflow.com} and \emph{GitHub.com} for different software engineering research purpose such as code recommendation, developer intent analysis, mining code repository, etc., but this is the first work with such a large set of user search data and the study identifies the popularity of these community-based Q\&A sites. Also recognizes that the necessity of community engagement for making a framework popular and easy to work with.}

\item \fh{Finally, with our proposed exception extraction model, exception type can be extracted without manual effort. This approach can also be extended for APIs, frameworks, etc. for user search behavior analysis.}

\end{itemize}

\vspace{-0.2cm}
\section{Threats to validity}
\label{sec:threat}
Our empirical study has some limitations that we would like to recognize:

Construct validity: Our metrics defined for success might not directly be identifiable with success in all scenarios. This is alleviated to a large degree by the fact that these are standard metrics ~\cite{Hu2011, Fox2005} defined and used in the IR~\cite{radlinski2005query, castillo2010, oj17-sigir} community for several years.

External Validity: There are three main external validity concerns. Temporal: Our results are obtained in the given time frame only. It is possible that the results might vary for a different time frame. Given our unintentional and 50 day time period selection, we hope to alleviate this problem. Geographic: Our data is based on search queries from United States with only English language queries. Analyzing differences in behavior across different locales, geographies and client form factors is an interesting and important topic but is out of scope for this work. Selection bias: The results could possibly be different should another GPSE be used. We think this is alleviated by the fact that we performed our analysis on a large sample of 5 billion search queries.

\vspace{-0.1cm}
\section{Conclusion}
\label{sec:conclusion}

\fh{In this paper, we have investigated for the first time how users search and find information about exceptions using web search. Our study identified one key implication that even with having more verbose exception message, exception search can suffer with less search success. Development community should look carefully at the reasons behind the low search success. Also, our study identifies the importance of community Q\&A sites for faster development and debugging. This information helps identify and motivate the importance of improving the documentation support for exceptions. Developer websites like StackOverflow and GitHub can also leverage the methodology and metrics proposed in this work for improving developer experience. Given the large body of work on software engineering recommendations~\cite{Holmes2005}, adding tool support in IDE's to have better suggestions for fixing exceptions would be of strong interest to the broader community. Additionally, in public forums, enhanced exception documentation or solution suggestions for most frequent, most search effort-intensive, or less successful exceptions can reduce developer effort. In the future, we plan to incorporate qualitative study to analyze exception search expectation from developer's point of view.}
\\
\textbf{Data sharing and availability:} The search data unfortunately cannot be shared publicly. This is due to legal laws and not due to independent choice. Search queries are very personal data and GDPR~\cite{gdpr} in Europe and equivalent privacy laws in other countries strictly govern the access to, usage and research that can be carried out on this data without specifically identifying an individual or groups of individuals. Interested researchers should contact us about the availability of similar data. Upon completion of the necessary legal steps and the legal paperwork, it may be possible to give access to similar search data for academic researchers.

%
\bibliographystyle{ACM-Reference-Format}
\bibliography{references}


\begin{thebibliography}{72}


\ifx \showCODEN    \undefined \def \showCODEN     #1{\unskip}     \fi
\ifx \showDOI      \undefined \def \showDOI       #1{#1}\fi
\ifx \showISBNx    \undefined \def \showISBNx     #1{\unskip}     \fi
\ifx \showISBNxiii \undefined \def \showISBNxiii  #1{\unskip}     \fi
\ifx \showISSN     \undefined \def \showISSN      #1{\unskip}     \fi
\ifx \showLCCN     \undefined \def \showLCCN      #1{\unskip}     \fi
\ifx \shownote     \undefined \def \shownote      #1{#1}          \fi
\ifx \showarticletitle \undefined \def \showarticletitle #1{#1}   \fi
\ifx \showURL      \undefined \def \showURL       {\relax}        \fi
\providecommand\bibfield[2]{#2}
\providecommand\bibinfo[2]{#2}
\providecommand\natexlab[1]{#1}
\providecommand\showeprint[2][]{arXiv:#2}

\bibitem[\protect\citeauthoryear{??}{cas}{2010}]%
        {castillo2010}
 \bibinfo{year}{2010}\natexlab{}.
\newblock \showarticletitle{The effects of query bursts on web search}. In
  \bibinfo{booktitle}{\emph{Proceedings - 2010 IEEE/WIC/ACM International
  Conference on Web Intelligence, WI 2010}}, Vol.~\bibinfo{volume}{1}.
  \bibinfo{pages}{374--381}.
\newblock
\showISBNx{9780769541914}
\urldef\tempurl%
\url{https://doi.org/10.1109/WI-IAT.2010.59}
\showDOI{\tempurl}


\bibitem[\protect\citeauthoryear{??}{csh}{2019}]%
        {csharpceptions}
 \bibinfo{year}{2019}\natexlab{}.
\newblock \bibinfo{title}{CSharp Exception List}.
\newblock
  \bibinfo{howpublished}{\url{https://powershellexplained.com/2017-04-07-all-dotnet-exception-list}}.
\newblock
\newblock
\shownote{Accessed: 2019-06-20.}


\bibitem[\protect\citeauthoryear{??}{jav}{2019}]%
        {javaexceptions}
 \bibinfo{year}{2019}\natexlab{}.
\newblock \bibinfo{title}{Java Exception List}.
\newblock
  \bibinfo{howpublished}{\url{https://programming.guide/java/list-of-java-exceptions.html}}.
\newblock
\newblock
\shownote{Accessed: 2019-06-20.}


\bibitem[\protect\citeauthoryear{??}{par}{2019}]%
        {parserator}
 \bibinfo{year}{2019}\natexlab{}.
\newblock \bibinfo{title}{Parserator}.
\newblock \bibinfo{howpublished}{\url{https://parserator.datamade.us/}}.
\newblock
\newblock
\shownote{Accessed: 2019-06-20.}


\bibitem[\protect\citeauthoryear{??}{pyt}{2019}]%
        {pythonexceptions}
 \bibinfo{year}{2019}\natexlab{}.
\newblock \bibinfo{title}{Python Exception List}.
\newblock
  \bibinfo{howpublished}{\url{https://docs.python.org/3/library/exceptions.html}}.
\newblock
\newblock
\shownote{Accessed: 2019-06-20.}


\bibitem[\protect\citeauthoryear{??}{kru}{2020}]%
        {krugle}
 \bibinfo{year}{2020}\natexlab{}.
\newblock \bibinfo{title}{Krugle}.
\newblock \bibinfo{howpublished}{\url{https://www.krugle.com/}}.
\newblock
\newblock
\shownote{Accessed: 2020-04-30.}


\bibitem[\protect\citeauthoryear{Ai, Dumais, Craswell, and Liebling}{Ai
  et~al\mbox{.}}{2017}]%
        {QAi2017}
\bibfield{author}{\bibinfo{person}{Qingyao Ai}, \bibinfo{person}{Susan~T.
  Dumais}, \bibinfo{person}{Nick Craswell}, {and} \bibinfo{person}{Dan
  Liebling}.} \bibinfo{year}{2017}\natexlab{}.
\newblock \showarticletitle{Characterizing Email Search Using Large-Scale
  Behavioral Logs and Surveys}. In \bibinfo{booktitle}{\emph{Proceedings of the
  26th International Conference on World Wide Web}} (Perth, Australia)
  \emph{(\bibinfo{series}{WWW '17})}. \bibinfo{publisher}{International World
  Wide Web Conferences Steering Committee}, \bibinfo{address}{Republic and
  Canton of Geneva, CHE}, \bibinfo{pages}{1511--1520}.
\newblock
\showISBNx{9781450349130}
\urldef\tempurl%
\url{https://doi.org/10.1145/3038912.3052615}
\showDOI{\tempurl}


\bibitem[\protect\citeauthoryear{Azzopardi, Kelly, and Brennan}{Azzopardi
  et~al\mbox{.}}{2013}]%
        {akb13-sigir}
\bibfield{author}{\bibinfo{person}{L. Azzopardi}, \bibinfo{person}{D. Kelly},
  {and} \bibinfo{person}{K. Brennan}.} \bibinfo{year}{2013}\natexlab{}.
\newblock \showarticletitle{How Query Cost Affects Search Behavior}.
  \bibinfo{pages}{23--32}.
\newblock


\bibitem[\protect\citeauthoryear{Azzopardi, White, Thomas, and
  Craswell}{Azzopardi et~al\mbox{.}}{2020}]%
        {Azzopardi2020}
\bibfield{author}{\bibinfo{person}{Leif Azzopardi}, \bibinfo{person}{Ryen~W.
  White}, \bibinfo{person}{Paul Thomas}, {and} \bibinfo{person}{Nick
  Craswell}.} \bibinfo{year}{2020}\natexlab{}.
\newblock \showarticletitle{Data-Driven Evaluation Metrics for Heterogeneous
  Search Engine Result Pages}. In \bibinfo{booktitle}{\emph{Proceedings of the
  2020 Conference on Human Information Interaction and Retrieval}} (Vancouver
  BC, Canada) \emph{(\bibinfo{series}{CHIIR '20})}.
  \bibinfo{publisher}{Association for Computing Machinery},
  \bibinfo{address}{New York, NY, USA}, \bibinfo{pages}{213--222}.
\newblock
\showISBNx{9781450368926}
\urldef\tempurl%
\url{https://doi.org/10.1145/3343413.3377959}
\showDOI{\tempurl}


\bibitem[\protect\citeauthoryear{Bajracharya, Ngo, Linstead, Dou, Rigor, Baldi,
  and Lopes}{Bajracharya et~al\mbox{.}}{2006}]%
        {bajracharya2006sourcerer}
\bibfield{author}{\bibinfo{person}{Sushil Bajracharya}, \bibinfo{person}{Trung
  Ngo}, \bibinfo{person}{Erik Linstead}, \bibinfo{person}{Yimeng Dou},
  \bibinfo{person}{Paul Rigor}, \bibinfo{person}{Pierre Baldi}, {and}
  \bibinfo{person}{Cristina Lopes}.} \bibinfo{year}{2006}\natexlab{}.
\newblock \showarticletitle{Sourcerer: a search engine for open source code
  supporting structure-based search}. In \bibinfo{booktitle}{\emph{Companion to
  the 21st ACM SIGPLAN symposium on Object-oriented programming systems,
  languages, and applications}}. ACM, \bibinfo{pages}{681--682}.
\newblock


\bibitem[\protect\citeauthoryear{Bajracharya and Lopes}{Bajracharya and
  Lopes}{2012}]%
        {Bajracharya2012}
\bibfield{author}{\bibinfo{person}{Sushil~Krishna Bajracharya} {and}
  \bibinfo{person}{Cristina~Videira Lopes}.} \bibinfo{year}{2012}\natexlab{}.
\newblock \showarticletitle{Analyzing and mining a code search engine usage
  log}.
\newblock \bibinfo{journal}{\emph{Empirical Software Engineering}}
  \bibinfo{volume}{17}, \bibinfo{number}{4} (\bibinfo{date}{01 Aug}
  \bibinfo{year}{2012}), \bibinfo{pages}{424--466}.
\newblock
\urldef\tempurl%
\url{https://doi.org/10.1007/s10664-010-9144-6}
\showDOI{\tempurl}


\bibitem[\protect\citeauthoryear{Bansal, Deligiannis, Maddila, and Rao}{Bansal
  et~al\mbox{.}}{2020}]%
        {bansal2020studying}
\bibfield{author}{\bibinfo{person}{Chetan Bansal}, \bibinfo{person}{Pantazis
  Deligiannis}, \bibinfo{person}{Chandra Maddila}, {and}
  \bibinfo{person}{Nikitha Rao}.} \bibinfo{year}{2020}\natexlab{}.
\newblock \showarticletitle{Studying Ransomware Attacks Using Web Search Logs}.
  In \bibinfo{booktitle}{\emph{Proceedings of the 43rd international ACM SIGIR
  conference on Research \& development in information retrieval}}.
\newblock
\urldef\tempurl%
\url{https://doi.org/10.1145/3397271.3401189}
\showDOI{\tempurl}


\bibitem[\protect\citeauthoryear{Bansal, Zimmermann, Awadallah, and
  Nagappan}{Bansal et~al\mbox{.}}{2019}]%
        {bansal2019usage}
\bibfield{author}{\bibinfo{person}{Chetan Bansal}, \bibinfo{person}{Thomas
  Zimmermann}, \bibinfo{person}{Ahmed~Hassan Awadallah}, {and}
  \bibinfo{person}{Nachiappan Nagappan}.} \bibinfo{year}{2019}\natexlab{}.
\newblock \showarticletitle{The Usage of Web Search for Software Engineering}.
\newblock \bibinfo{journal}{\emph{arXiv preprint arXiv:1912.09519}}
  (\bibinfo{year}{2019}).
\newblock


\bibitem[\protect\citeauthoryear{Carlson, Gaffney, and Vasile}{Carlson
  et~al\mbox{.}}{2009}]%
        {carlson2009learning}
\bibfield{author}{\bibinfo{person}{Andrew Carlson}, \bibinfo{person}{Scott
  Gaffney}, {and} \bibinfo{person}{Flavian Vasile}.}
  \bibinfo{year}{2009}\natexlab{}.
\newblock \showarticletitle{Learning a Named Entity Tagger from Gazetteers with
  the Partial Perceptron.}. In \bibinfo{booktitle}{\emph{AAAI Spring Symposium:
  Learning by Reading and Learning to Read}}. \bibinfo{pages}{7--13}.
\newblock


\bibitem[\protect\citeauthoryear{Chancellor and Counts}{Chancellor and
  Counts}{2018}]%
        {chancellor2018measuring}
\bibfield{author}{\bibinfo{person}{Stevie Chancellor} {and}
  \bibinfo{person}{Scott Counts}.} \bibinfo{year}{2018}\natexlab{}.
\newblock \showarticletitle{Measuring employment demand using internet search
  data}. In \bibinfo{booktitle}{\emph{Proceedings of the 2018 CHI Conference on
  Human Factors in Computing Systems}}. ACM, \bibinfo{pages}{122}.
\newblock


\bibitem[\protect\citeauthoryear{Chen, Lee, Xie, Yang, Lasecki, and Oney}{Chen
  et~al\mbox{.}}{2017}]%
        {Chen2017}
\bibfield{author}{\bibinfo{person}{Yan Chen}, \bibinfo{person}{Sang~Won Lee},
  \bibinfo{person}{Yin Xie}, \bibinfo{person}{YiWei Yang},
  \bibinfo{person}{Walter~S. Lasecki}, {and} \bibinfo{person}{Steve Oney}.}
  \bibinfo{year}{2017}\natexlab{}.
\newblock \showarticletitle{Codeon: On-Demand Software Development Assistance}.
  In \bibinfo{booktitle}{\emph{Proceedings of the 2017 CHI Conference on Human
  Factors in Computing Systems}} (Denver, Colorado, USA)
  \emph{(\bibinfo{series}{CHI '17})}. \bibinfo{publisher}{Association for
  Computing Machinery}, \bibinfo{address}{New York, NY, USA},
  \bibinfo{pages}{6220--6231}.
\newblock
\showISBNx{9781450346559}
\urldef\tempurl%
\url{https://doi.org/10.1145/3025453.3025972}
\showDOI{\tempurl}


\bibitem[\protect\citeauthoryear{Collier, Nobata, and Tsujii}{Collier
  et~al\mbox{.}}{2000}]%
        {collier2000extracting}
\bibfield{author}{\bibinfo{person}{Nigel Collier}, \bibinfo{person}{Chikashi
  Nobata}, {and} \bibinfo{person}{Jun-ichi Tsujii}.}
  \bibinfo{year}{2000}\natexlab{}.
\newblock \showarticletitle{Extracting the names of genes and gene products
  with a hidden Markov model}. In \bibinfo{booktitle}{\emph{Proceedings of the
  18th conference on Computational linguistics-Volume 1}}. Association for
  Computational Linguistics, \bibinfo{pages}{201--207}.
\newblock


\bibitem[\protect\citeauthoryear{{Corley}, {Lois}, and {Quezada}}{{Corley}
  et~al\mbox{.}}{2015}]%
        {Corley2015}
\bibfield{author}{\bibinfo{person}{C.~S. {Corley}}, \bibinfo{person}{F.
  {Lois}}, {and} \bibinfo{person}{S. {Quezada}}.}
  \bibinfo{year}{2015}\natexlab{}.
\newblock \showarticletitle{Web usage patterns of developers}. In
  \bibinfo{booktitle}{\emph{2015 IEEE International Conference on Software
  Maintenance and Evolution (ICSME)}}. \bibinfo{pages}{381--390}.
\newblock
\urldef\tempurl%
\url{https://doi.org/10.1109/ICSM.2015.7332489}
\showDOI{\tempurl}


\bibitem[\protect\citeauthoryear{Durieux, Abreu, Monperrus, and
  Bissyand{\'{e}}}{Durieux et~al\mbox{.}}{2019}]%
        {Durieux2019}
\bibfield{author}{\bibinfo{person}{Thomas Durieux}, \bibinfo{person}{Rui
  Abreu}, \bibinfo{person}{Martin Monperrus}, {and}
  \bibinfo{person}{Tegawend{\'{e}}~F. Bissyand{\'{e}}}.}
  \bibinfo{year}{2019}\natexlab{}.
\newblock \showarticletitle{Interviewing the Most Successful Bot on GitHub: Dr
  Travis {CI} on 35+ Million of its Jobs}.
\newblock \bibinfo{journal}{\emph{CoRR}}  \bibinfo{volume}{abs/1904.09416}
  (\bibinfo{year}{2019}).
\newblock
\urldef\tempurl%
\url{http://arxiv.org/abs/1904.09416}
\showURL{%
\tempurl}


\bibitem[\protect\citeauthoryear{Fayolle, Moreau, Raymond, Gravier, and
  Gros}{Fayolle et~al\mbox{.}}{2010}]%
        {fayolle2010crf}
\bibfield{author}{\bibinfo{person}{Julien Fayolle}, \bibinfo{person}{Fabienne
  Moreau}, \bibinfo{person}{Christian Raymond}, \bibinfo{person}{Guillaume
  Gravier}, {and} \bibinfo{person}{Patrick Gros}.}
  \bibinfo{year}{2010}\natexlab{}.
\newblock \showarticletitle{CRF-based combination of contextual features to
  improve a posteriori word-level confidence measures}. In
  \bibinfo{booktitle}{\emph{Eleventh Annual Conference of the International
  Speech Communication Association}}.
\newblock


\bibitem[\protect\citeauthoryear{Fox, Karnawat, Mydland, Dumais, and White}{Fox
  et~al\mbox{.}}{2005}]%
        {Fox2005}
\bibfield{author}{\bibinfo{person}{Steve Fox}, \bibinfo{person}{Kuldeep
  Karnawat}, \bibinfo{person}{Mark Mydland}, \bibinfo{person}{Susan Dumais},
  {and} \bibinfo{person}{Thomas White}.} \bibinfo{year}{2005}\natexlab{}.
\newblock \showarticletitle{Evaluating Implicit Measures to Improve Web
  Search}.
\newblock \bibinfo{journal}{\emph{ACM Trans. Inf. Syst.}} \bibinfo{volume}{23},
  \bibinfo{number}{2} (\bibinfo{date}{April} \bibinfo{year}{2005}).
\newblock


\bibitem[\protect\citeauthoryear{Friedman, Hastie, and Tibshirani}{Friedman
  et~al\mbox{.}}{2010}]%
        {l1l2}
\bibfield{author}{\bibinfo{person}{Jerome Friedman}, \bibinfo{person}{Trevor
  Hastie}, {and} \bibinfo{person}{Rob Tibshirani}.}
  \bibinfo{year}{2010}\natexlab{}.
\newblock \showarticletitle{Regularization paths for generalized linear models
  via coordinate descent}.
\newblock \bibinfo{journal}{\emph{Journal of statistical software}}
  \bibinfo{volume}{33}, \bibinfo{number}{1} (\bibinfo{year}{2010}),
  \bibinfo{pages}{1}.
\newblock


\bibitem[\protect\citeauthoryear{gdpr.eu}{gdpr.eu}{2019}]%
        {gdpr}
\bibfield{author}{\bibinfo{person}{gdpr.eu}.} \bibinfo{year}{2019}\natexlab{}.
\newblock \bibinfo{title}{{General Data Protection Regulation}}.
\newblock \bibinfo{howpublished}{\url{https://gdpr-info.eu/}}.
\newblock


\bibitem[\protect\citeauthoryear{Haiduc, Bavota, Marcus, Oliveto, De~Lucia, and
  Menzies}{Haiduc et~al\mbox{.}}{2013}]%
        {Haiduc2013}
\bibfield{author}{\bibinfo{person}{Sonia Haiduc}, \bibinfo{person}{Gabriele
  Bavota}, \bibinfo{person}{Andrian Marcus}, \bibinfo{person}{Rocco Oliveto},
  \bibinfo{person}{Andrea De~Lucia}, {and} \bibinfo{person}{Tim Menzies}.}
  \bibinfo{year}{2013}\natexlab{}.
\newblock \showarticletitle{Automatic Query Reformulations for Text Retrieval
  in Software Engineering}. In \bibinfo{booktitle}{\emph{Proceedings of the
  2013 International Conference on Software Engineering}} (San Francisco, CA,
  USA) \emph{(\bibinfo{series}{ICSE '13})}. \bibinfo{publisher}{IEEE Press},
  \bibinfo{address}{Piscataway, NJ, USA}, \bibinfo{pages}{842--851}.
\newblock
\showISBNx{978-1-4673-3076-3}
\urldef\tempurl%
\url{http://dl.acm.org/citation.cfm?id=2486788.2486898}
\showURL{%
\tempurl}


\bibitem[\protect\citeauthoryear{Hassan, Fahmy, and Hassan}{Hassan
  et~al\mbox{.}}{2007}]%
        {hassanawadallah2007improving}
\bibfield{author}{\bibinfo{person}{Ahmed Hassan}, \bibinfo{person}{Haytham
  Fahmy}, {and} \bibinfo{person}{Hany Hassan}.}
  \bibinfo{year}{2007}\natexlab{}.
\newblock \showarticletitle{Improving named entity translation by exploiting
  comparable and parallel corpora}.
\newblock \bibinfo{journal}{\emph{ACQUISITION AND MANAGEMENT OF MULTILINGUAL
  LEXICONS}} (\bibinfo{year}{2007}), \bibinfo{pages}{35}.
\newblock


\bibitem[\protect\citeauthoryear{Hassan, Jones, and Klinkner}{Hassan
  et~al\mbox{.}}{2010}]%
        {Hassan:2010:BDU:1718487.1718515}
\bibfield{author}{\bibinfo{person}{Ahmed Hassan}, \bibinfo{person}{Rosie
  Jones}, {and} \bibinfo{person}{Kristina~Lisa Klinkner}.}
  \bibinfo{year}{2010}\natexlab{}.
\newblock \showarticletitle{Beyond DCG: User Behavior As a Predictor of a
  Successful Search}. In \bibinfo{booktitle}{\emph{Proceedings of the Third ACM
  International Conference on Web Search and Data Mining}} (New York, New York,
  USA) \emph{(\bibinfo{series}{WSDM '10})}. \bibinfo{publisher}{ACM},
  \bibinfo{address}{New York, NY, USA}, \bibinfo{pages}{221--230}.
\newblock
\showISBNx{978-1-60558-889-6}
\urldef\tempurl%
\url{https://doi.org/10.1145/1718487.1718515}
\showDOI{\tempurl}


\bibitem[\protect\citeauthoryear{{Hassan} and {Xiaoyin Wang}}{{Hassan} and
  {Xiaoyin Wang}}{2017}]%
        {foyzul2017}
\bibfield{author}{\bibinfo{person}{F. {Hassan}} {and} \bibinfo{person}{{Xiaoyin
  Wang}}.} \bibinfo{year}{2017}\natexlab{}.
\newblock \showarticletitle{Mining Readme Files to Support Automatic Building
  of Java Projects in Software Repositories}. In \bibinfo{booktitle}{\emph{2017
  IEEE/ACM 39th International Conference on Software Engineering Companion
  (ICSE-C)}}. \bibinfo{pages}{277--279}.
\newblock
\urldef\tempurl%
\url{https://doi.org/10.1109/ICSE-C.2017.114}
\showDOI{\tempurl}


\bibitem[\protect\citeauthoryear{Holmes, Walker, and Murphy}{Holmes
  et~al\mbox{.}}{2005}]%
        {Holmes2005}
\bibfield{author}{\bibinfo{person}{Reid Holmes}, \bibinfo{person}{Robert~J.
  Walker}, {and} \bibinfo{person}{Gail~C. Murphy}.}
  \bibinfo{year}{2005}\natexlab{}.
\newblock \showarticletitle{Strathcona Example Recommendation Tool}. In
  \bibinfo{booktitle}{\emph{Proceedings of the 10th European Software
  Engineering Conference Held Jointly with 13th ACM SIGSOFT International
  Symposium on Foundations of Software Engineering}} (Lisbon, Portugal)
  \emph{(\bibinfo{series}{ESEC/FSE-13})}. \bibinfo{publisher}{ACM},
  \bibinfo{address}{New York, NY, USA}, \bibinfo{pages}{237--240}.
\newblock
\showISBNx{1-59593-014-0}
\urldef\tempurl%
\url{https://doi.org/10.1145/1081706.1081744}
\showDOI{\tempurl}


\bibitem[\protect\citeauthoryear{Hu, Zhang, Chen, Wang, and Yang}{Hu
  et~al\mbox{.}}{2011}]%
        {Hu2011}
\bibfield{author}{\bibinfo{person}{Botao Hu}, \bibinfo{person}{Yuchen Zhang},
  \bibinfo{person}{Weizhu Chen}, \bibinfo{person}{Gang Wang}, {and}
  \bibinfo{person}{Qiang Yang}.} \bibinfo{year}{2011}\natexlab{}.
\newblock \showarticletitle{Characterizing Search Intent Diversity into Click
  Models}. In \bibinfo{booktitle}{\emph{Proceedings of the 20th International
  Conference on World Wide Web}} (Hyderabad, India) \emph{(\bibinfo{series}{WWW
  '11})}. \bibinfo{publisher}{ACM}, \bibinfo{address}{New York, NY, USA},
  \bibinfo{pages}{17--26}.
\newblock
\showISBNx{978-1-4503-0632-4}
\urldef\tempurl%
\url{https://doi.org/10.1145/1963405.1963412}
\showDOI{\tempurl}


\bibitem[\protect\citeauthoryear{Jhaver, Cranshaw, and Counts}{Jhaver
  et~al\mbox{.}}{2019}]%
        {jhaver2019measuring}
\bibfield{author}{\bibinfo{person}{Shagun Jhaver}, \bibinfo{person}{Justin
  Cranshaw}, {and} \bibinfo{person}{Scott Counts}.}
  \bibinfo{year}{2019}\natexlab{}.
\newblock \showarticletitle{Measuring Professional Skill Development in US
  Cities Using Internet Search Queries}. In
  \bibinfo{booktitle}{\emph{Proceedings of the International AAAI Conference on
  Web and Social Media}}, Vol.~\bibinfo{volume}{13}. \bibinfo{pages}{267--277}.
\newblock


\bibitem[\protect\citeauthoryear{{Jia}, {Li}, {Yu}, {Liao}, {Wang}, {Liu}, and
  {Liu}}{{Jia} et~al\mbox{.}}{2019}]%
        {ZJia2019}
\bibfield{author}{\bibinfo{person}{Z. {Jia}}, \bibinfo{person}{S. {Li}},
  \bibinfo{person}{T. {Yu}}, \bibinfo{person}{X. {Liao}}, \bibinfo{person}{J.
  {Wang}}, \bibinfo{person}{X. {Liu}}, {and} \bibinfo{person}{Y. {Liu}}.}
  \bibinfo{year}{2019}\natexlab{}.
\newblock \showarticletitle{Detecting Error-Handling Bugs without Error
  Specification Input}. In \bibinfo{booktitle}{\emph{2019 34th IEEE/ACM
  International Conference on Automated Software Engineering (ASE)}}.
  \bibinfo{pages}{213--225}.
\newblock


\bibitem[\protect\citeauthoryear{Jiang, Hassan~Awadallah, Shi, and White}{Jiang
  et~al\mbox{.}}{2015}]%
        {Jiang2015}
\bibfield{author}{\bibinfo{person}{Jiepu Jiang}, \bibinfo{person}{Ahmed
  Hassan~Awadallah}, \bibinfo{person}{Xiaolin Shi}, {and}
  \bibinfo{person}{Ryen~W. White}.} \bibinfo{year}{2015}\natexlab{}.
\newblock \showarticletitle{Understanding and Predicting Graded Search
  Satisfaction}. In \bibinfo{booktitle}{\emph{Proceedings of the Eighth ACM
  International Conference on Web Search and Data Mining}} (Shanghai, China)
  \emph{(\bibinfo{series}{WSDM '15})}. \bibinfo{publisher}{ACM},
  \bibinfo{address}{New York, NY, USA}, \bibinfo{pages}{57--66}.
\newblock
\showISBNx{978-1-4503-3317-7}
\urldef\tempurl%
\url{https://doi.org/10.1145/2684822.2685319}
\showDOI{\tempurl}


\bibitem[\protect\citeauthoryear{Jonnalagadda, Cohen, Wu, Liu, and
  Gonzalez}{Jonnalagadda et~al\mbox{.}}{2013}]%
        {jonnalagadda2013using}
\bibfield{author}{\bibinfo{person}{Siddhartha Jonnalagadda},
  \bibinfo{person}{Trevor Cohen}, \bibinfo{person}{Stephen Wu},
  \bibinfo{person}{Hongfang Liu}, {and} \bibinfo{person}{Graciela Gonzalez}.}
  \bibinfo{year}{2013}\natexlab{}.
\newblock \showarticletitle{Using empirically constructed lexical resources for
  named entity recognition}.
\newblock \bibinfo{journal}{\emph{Biomedical informatics insights}}
  \bibinfo{volume}{6} (\bibinfo{year}{2013}), \bibinfo{pages}{BII--S11664}.
\newblock


\bibitem[\protect\citeauthoryear{Kim, Kim, Bissyand\'{e}, Choi, Li, Klein, and
  Traon}{Kim et~al\mbox{.}}{2018}]%
        {Kim2018}
\bibfield{author}{\bibinfo{person}{Kisub Kim}, \bibinfo{person}{Dongsun Kim},
  \bibinfo{person}{Tegawend\'{e}~F. Bissyand\'{e}}, \bibinfo{person}{Eunjong
  Choi}, \bibinfo{person}{Li Li}, \bibinfo{person}{Jacques Klein}, {and}
  \bibinfo{person}{Yves~Le Traon}.} \bibinfo{year}{2018}\natexlab{}.
\newblock \showarticletitle{FaCoY: A Code-to-Code Search Engine}. In
  \bibinfo{booktitle}{\emph{Proceedings of the 40th International Conference on
  Software Engineering}} (Gothenburg, Sweden) \emph{(\bibinfo{series}{ICSE
  '18})}. \bibinfo{publisher}{Association for Computing Machinery},
  \bibinfo{address}{New York, NY, USA}, \bibinfo{pages}{946--957}.
\newblock
\showISBNx{9781450356381}
\urldef\tempurl%
\url{https://doi.org/10.1145/3180155.3180187}
\showDOI{\tempurl}


\bibitem[\protect\citeauthoryear{Lafferty, McCallum, and Pereira}{Lafferty
  et~al\mbox{.}}{2001}]%
        {Lafferty2001}
\bibfield{author}{\bibinfo{person}{John~D. Lafferty}, \bibinfo{person}{Andrew
  McCallum}, {and} \bibinfo{person}{Fernando C.~N. Pereira}.}
  \bibinfo{year}{2001}\natexlab{}.
\newblock \showarticletitle{Conditional Random Fields: Probabilistic Models for
  Segmenting and Labeling Sequence Data}. In
  \bibinfo{booktitle}{\emph{Proceedings of the Eighteenth International
  Conference on Machine Learning}} \emph{(\bibinfo{series}{ICML '01})}.
  \bibinfo{publisher}{Morgan Kaufmann Publishers Inc.}, \bibinfo{address}{San
  Francisco, CA, USA}, \bibinfo{pages}{282--289}.
\newblock
\showISBNx{1-55860-778-1}
\urldef\tempurl%
\url{http://dl.acm.org/citation.cfm?id=645530.655813}
\showURL{%
\tempurl}


\bibitem[\protect\citeauthoryear{Liu, Liu, and Yan}{Liu et~al\mbox{.}}{2018}]%
        {Liu2018}
\bibfield{author}{\bibinfo{person}{Chang Liu}, \bibinfo{person}{Jingjing Liu},
  {and} \bibinfo{person}{Zengwang Yan}.} \bibinfo{year}{2018}\natexlab{}.
\newblock \showarticletitle{Personalizing Information Retrieval Using Search
  Behaviors and Time Constraints}. In \bibinfo{booktitle}{\emph{Proceedings of
  the 2018 Conference on Human Information Interaction and Retrieval}} (New
  Brunswick, NJ, USA) \emph{(\bibinfo{series}{CHIIR '18})}.
  \bibinfo{publisher}{Association for Computing Machinery},
  \bibinfo{address}{New York, NY, USA}, \bibinfo{pages}{261--264}.
\newblock
\showISBNx{9781450349253}
\urldef\tempurl%
\url{https://doi.org/10.1145/3176349.3176878}
\showDOI{\tempurl}


\bibitem[\protect\citeauthoryear{Liu and Nocedal}{Liu and Nocedal}{1989}]%
        {lbgs1989}
\bibfield{author}{\bibinfo{person}{Dong~C. Liu} {and} \bibinfo{person}{Jorge
  Nocedal}.} \bibinfo{year}{1989}\natexlab{}.
\newblock \showarticletitle{On the Limited Memory BFGS Method for Large Scale
  Optimization}.
\newblock \bibinfo{journal}{\emph{Math. Program.}} \bibinfo{volume}{45},
  \bibinfo{number}{1-3} (\bibinfo{date}{Aug.} \bibinfo{year}{1989}),
  \bibinfo{pages}{503--528}.
\newblock
\showISSN{0025-5610}
\urldef\tempurl%
\url{http://dl.acm.org/citation.cfm?id=3112655.3112866}
\showURL{%
\tempurl}


\bibitem[\protect\citeauthoryear{Liu, Zhang, Wei, and Zhou}{Liu
  et~al\mbox{.}}{2011}]%
        {Liu2011}
\bibfield{author}{\bibinfo{person}{Xiaohua Liu}, \bibinfo{person}{Shaodian
  Zhang}, \bibinfo{person}{Furu Wei}, {and} \bibinfo{person}{Ming Zhou}.}
  \bibinfo{year}{2011}\natexlab{}.
\newblock \showarticletitle{Recognizing Named Entities in Tweets}. In
  \bibinfo{booktitle}{\emph{Proceedings of the 49th Annual Meeting of the
  Association for Computational Linguistics: Human Language Technologies -
  Volume 1}} (Portland, Oregon) \emph{(\bibinfo{series}{HLT '11})}.
  \bibinfo{publisher}{Association for Computational Linguistics},
  \bibinfo{address}{Stroudsburg, PA, USA}, \bibinfo{pages}{359--367}.
\newblock
\showISBNx{978-1-932432-87-9}
\urldef\tempurl%
\url{http://dl.acm.org/citation.cfm?id=2002472.2002519}
\showURL{%
\tempurl}


\bibitem[\protect\citeauthoryear{Lu, Zhang, and Ma}{Lu et~al\mbox{.}}{2018}]%
        {Lu18}
\bibfield{author}{\bibinfo{person}{Hongyu Lu}, \bibinfo{person}{Min Zhang},
  {and} \bibinfo{person}{Shaoping Ma}.} \bibinfo{year}{2018}\natexlab{}.
\newblock \showarticletitle{Between Clicks and Satisfaction: Study on
  Multi-Phase User Preferences and Satisfaction for Online News Reading}. In
  \bibinfo{booktitle}{\emph{The 41st International ACM SIGIR Conference on
  Research and Development in Information Retrieval}} (Ann Arbor, MI, USA)
  \emph{(\bibinfo{series}{SIGIR '18})}. \bibinfo{publisher}{Association for
  Computing Machinery}, \bibinfo{address}{New York, NY, USA},
  \bibinfo{pages}{435--444}.
\newblock
\showISBNx{9781450356572}
\urldef\tempurl%
\url{https://doi.org/10.1145/3209978.3210007}
\showDOI{\tempurl}


\bibitem[\protect\citeauthoryear{Luo, Zhang, Li, Liu, Zhang, Ma, and Yang}{Luo
  et~al\mbox{.}}{2016}]%
        {Luo16}
\bibfield{author}{\bibinfo{person}{Cheng Luo}, \bibinfo{person}{Fan Zhang},
  \bibinfo{person}{Xue Li}, \bibinfo{person}{Yiqun Liu}, \bibinfo{person}{Min
  Zhang}, \bibinfo{person}{Shaoping Ma}, {and} \bibinfo{person}{Delin Yang}.}
  \bibinfo{year}{2016}\natexlab{}.
\newblock \showarticletitle{Manipulating Time Perception of Web Search Users}.
  In \bibinfo{booktitle}{\emph{Proceedings of the 2016 ACM on Conference on
  Human Information Interaction and Retrieval}} (Carrboro, North Carolina, USA)
  \emph{(\bibinfo{series}{CHIIR '16})}. \bibinfo{publisher}{Association for
  Computing Machinery}, \bibinfo{address}{New York, NY, USA},
  \bibinfo{pages}{293--296}.
\newblock
\showISBNx{9781450337519}
\urldef\tempurl%
\url{https://doi.org/10.1145/2854946.2854994}
\showDOI{\tempurl}


\bibitem[\protect\citeauthoryear{Ma and Hovy}{Ma and Hovy}{2016}]%
        {ma2016end}
\bibfield{author}{\bibinfo{person}{Xuezhe Ma} {and} \bibinfo{person}{Eduard
  Hovy}.} \bibinfo{year}{2016}\natexlab{}.
\newblock \showarticletitle{End-to-end sequence labeling via bi-directional
  lstm-cnns-crf}.
\newblock \bibinfo{journal}{\emph{arXiv preprint arXiv:1603.01354}}
  (\bibinfo{year}{2016}).
\newblock


\bibitem[\protect\citeauthoryear{Mao, Liu, Zhou, Nie, Song, Zhang, Ma, Sun, and
  Luo}{Mao et~al\mbox{.}}{2016}]%
        {Mao2016}
\bibfield{author}{\bibinfo{person}{Jiaxin Mao}, \bibinfo{person}{Yiqun Liu},
  \bibinfo{person}{Ke Zhou}, \bibinfo{person}{Jian-Yun Nie},
  \bibinfo{person}{Jingtao Song}, \bibinfo{person}{Min Zhang},
  \bibinfo{person}{Shaoping Ma}, \bibinfo{person}{Jiashen Sun}, {and}
  \bibinfo{person}{Hengliang Luo}.} \bibinfo{year}{2016}\natexlab{}.
\newblock \showarticletitle{When Does Relevance Mean Usefulness and User
  Satisfaction in Web Search?}. In \bibinfo{booktitle}{\emph{Proceedings of the
  39th International ACM SIGIR Conference on Research and Development in
  Information Retrieval}} (Pisa, Italy) \emph{(\bibinfo{series}{SIGIR '16})}.
  \bibinfo{publisher}{Association for Computing Machinery},
  \bibinfo{address}{New York, NY, USA}, \bibinfo{pages}{463--472}.
\newblock
\showISBNx{9781450340694}
\urldef\tempurl%
\url{https://doi.org/10.1145/2911451.2911507}
\showDOI{\tempurl}


\bibitem[\protect\citeauthoryear{Martie, Hoek, and Kwak}{Martie
  et~al\mbox{.}}{2017}]%
        {Martie2017}
\bibfield{author}{\bibinfo{person}{Lee Martie}, \bibinfo{person}{Andr{\'e}
  van~der Hoek}, {and} \bibinfo{person}{Thomas Kwak}.}
  \bibinfo{year}{2017}\natexlab{}.
\newblock \showarticletitle{Understanding the Impact of Support for Iteration
  on Code Search}. In \bibinfo{booktitle}{\emph{Proceedings of the 2017 11th
  Joint Meeting on Foundations of Software Engineering}} (Paderborn, Germany)
  \emph{(\bibinfo{series}{ESEC/FSE 2017})}. \bibinfo{publisher}{ACM},
  \bibinfo{address}{New York, NY, USA}, \bibinfo{pages}{774--785}.
\newblock
\showISBNx{978-1-4503-5105-8}
\urldef\tempurl%
\url{https://doi.org/10.1145/3106237.3106293}
\showDOI{\tempurl}


\bibitem[\protect\citeauthoryear{Mehrotra, Anderson, Diaz, Sharma, Wallach, and
  Yilmaz}{Mehrotra et~al\mbox{.}}{2017}]%
        {Mehrotra2017}
\bibfield{author}{\bibinfo{person}{Rishabh Mehrotra}, \bibinfo{person}{Ashton
  Anderson}, \bibinfo{person}{Fernando Diaz}, \bibinfo{person}{Amit Sharma},
  \bibinfo{person}{Hanna Wallach}, {and} \bibinfo{person}{Emine Yilmaz}.}
  \bibinfo{year}{2017}\natexlab{}.
\newblock \showarticletitle{Auditing Search Engines for Differential
  Satisfaction Across Demographics}. In \bibinfo{booktitle}{\emph{Proceedings
  of the 26th International Conference on World Wide Web Companion}}
  \emph{(\bibinfo{series}{WWW '17 Companion})}.
\newblock


\bibitem[\protect\citeauthoryear{Meng, Nagy, Yao, Zhuang, and Argoty}{Meng
  et~al\mbox{.}}{2018}]%
        {Meng2018}
\bibfield{author}{\bibinfo{person}{Na Meng}, \bibinfo{person}{Stefan Nagy},
  \bibinfo{person}{Danfeng~(Daphne) Yao}, \bibinfo{person}{Wenjie Zhuang},
  {and} \bibinfo{person}{Gustavo~Arango Argoty}.}
  \bibinfo{year}{2018}\natexlab{}.
\newblock \showarticletitle{Secure Coding Practices in Java: Challenges and
  Vulnerabilities}. In \bibinfo{booktitle}{\emph{Proceedings of the 40th
  International Conference on Software Engineering}} (Gothenburg, Sweden)
  \emph{(\bibinfo{series}{ICSE '18})}. \bibinfo{publisher}{Association for
  Computing Machinery}, \bibinfo{address}{New York, NY, USA},
  \bibinfo{pages}{372--383}.
\newblock
\showISBNx{9781450356381}
\urldef\tempurl%
\url{https://doi.org/10.1145/3180155.3180201}
\showDOI{\tempurl}


\bibitem[\protect\citeauthoryear{Middleton and Krivcovs}{Middleton and
  Krivcovs}{2016}]%
        {Middleton2016}
\bibfield{author}{\bibinfo{person}{Stuart~E. Middleton} {and}
  \bibinfo{person}{Vadims Krivcovs}.} \bibinfo{year}{2016}\natexlab{}.
\newblock \showarticletitle{Geoparsing and Geosemantics for Social Media:
  Spatiotemporal Grounding of Content Propagating Rumors to Support Trust and
  Veracity Analysis During Breaking News}.
\newblock \bibinfo{journal}{\emph{ACM Trans. Inf. Syst.}} \bibinfo{volume}{34},
  \bibinfo{number}{3}, Article \bibinfo{articleno}{16} (\bibinfo{date}{April}
  \bibinfo{year}{2016}), \bibinfo{numpages}{26}~pages.
\newblock
\showISSN{1046-8188}
\urldef\tempurl%
\url{https://doi.org/10.1145/2842604}
\showDOI{\tempurl}


\bibitem[\protect\citeauthoryear{Mozilla.org}{Mozilla.org}{2019}]%
        {AcceptLanguage-HTTP}
\bibfield{author}{\bibinfo{person}{Mozilla.org}.}
  \bibinfo{year}{2019}\natexlab{}.
\newblock \bibinfo{title}{{Accept-Language HTTP}}.
\newblock
  \bibinfo{howpublished}{\url{https://developer.mozilla.org/en-US/docs/Web/HTTP/Headers/Accept-Language}}.
\newblock


\bibitem[\protect\citeauthoryear{{Nanz} and {Furia}}{{Nanz} and
  {Furia}}{2015}]%
        {Nanz2015}
\bibfield{author}{\bibinfo{person}{S. {Nanz}} {and} \bibinfo{person}{C.~A.
  {Furia}}.} \bibinfo{year}{2015}\natexlab{}.
\newblock \showarticletitle{A Comparative Study of Programming Languages in
  Rosetta Code}. In \bibinfo{booktitle}{\emph{2015 IEEE/ACM 37th IEEE
  International Conference on Software Engineering}}, Vol.~\bibinfo{volume}{1}.
  \bibinfo{pages}{778--788}.
\newblock
\showISSN{0270-5257}
\urldef\tempurl%
\url{https://doi.org/10.1109/ICSE.2015.90}
\showDOI{\tempurl}


\bibitem[\protect\citeauthoryear{Nasehi, Sillito, Maurer, and Burns}{Nasehi
  et~al\mbox{.}}{2012}]%
        {nasehi2012makes}
\bibfield{author}{\bibinfo{person}{Seyed~Mehdi Nasehi},
  \bibinfo{person}{Jonathan Sillito}, \bibinfo{person}{Frank Maurer}, {and}
  \bibinfo{person}{Chris Burns}.} \bibinfo{year}{2012}\natexlab{}.
\newblock \showarticletitle{What makes a good code example?: A study of
  programming Q\&A in StackOverflow}. In \bibinfo{booktitle}{\emph{2012 28th
  IEEE International Conference on Software Maintenance (ICSM)}}. IEEE,
  \bibinfo{pages}{25--34}.
\newblock


\bibitem[\protect\citeauthoryear{{Nguyen}, {Rigby}, {Nguyen}, {Palani},
  {Karanfil}, and {Nguyen}}{{Nguyen} et~al\mbox{.}}{2018}]%
        {Nguyen2018}
\bibfield{author}{\bibinfo{person}{A.~T. {Nguyen}}, \bibinfo{person}{P.~C.
  {Rigby}}, \bibinfo{person}{T. {Nguyen}}, \bibinfo{person}{D. {Palani}},
  \bibinfo{person}{M. {Karanfil}}, {and} \bibinfo{person}{T.~N. {Nguyen}}.}
  \bibinfo{year}{2018}\natexlab{}.
\newblock \showarticletitle{Statistical Translation of English Texts to API
  Code Templates}. In \bibinfo{booktitle}{\emph{2018 IEEE International
  Conference on Software Maintenance and Evolution (ICSME)}}.
  \bibinfo{pages}{194--205}.
\newblock
\showISSN{2576-3148}
\urldef\tempurl%
\url{https://doi.org/10.1109/ICSME.2018.00029}
\showDOI{\tempurl}


\bibitem[\protect\citeauthoryear{Okazaki}{Okazaki}{2007}]%
        {CRFsuite}
\bibfield{author}{\bibinfo{person}{Naoaki Okazaki}.}
  \bibinfo{year}{2007}\natexlab{}.
\newblock \bibinfo{title}{CRFsuite: a fast implementation of Conditional Random
  Fields (CRFs)}.
\newblock
\newblock
\urldef\tempurl%
\url{http://www.chokkan.org/software/crfsuite/}
\showURL{%
\tempurl}


\bibitem[\protect\citeauthoryear{Ong, J\"{a}rvelin, Sanderson, and Scholer}{Ong
  et~al\mbox{.}}{2017}]%
        {oj17-sigir}
\bibfield{author}{\bibinfo{person}{K. Ong}, \bibinfo{person}{K. J\"{a}rvelin},
  \bibinfo{person}{M. Sanderson}, {and} \bibinfo{person}{F. Scholer}.}
  \bibinfo{year}{2017}\natexlab{}.
\newblock \showarticletitle{Using Information Scent to Understand Mobile and
  Desktop Web Search Behavior}. \bibinfo{pages}{295--304}.
\newblock


\bibitem[\protect\citeauthoryear{Paparrizos, White, and Horvitz}{Paparrizos
  et~al\mbox{.}}{2016}]%
        {paparrizos2016detecting}
\bibfield{author}{\bibinfo{person}{John Paparrizos}, \bibinfo{person}{Ryen~W
  White}, {and} \bibinfo{person}{Eric Horvitz}.}
  \bibinfo{year}{2016}\natexlab{}.
\newblock \showarticletitle{Detecting devastating diseases in search logs}. In
  \bibinfo{booktitle}{\emph{Proceedings of the 22nd ACM SIGKDD International
  Conference on Knowledge Discovery and Data Mining}}. ACM,
  \bibinfo{pages}{559--568}.
\newblock


\bibitem[\protect\citeauthoryear{Rabiner}{Rabiner}{1990}]%
        {Rabiner1990}
\bibfield{author}{\bibinfo{person}{Lawrence~R. Rabiner}.}
  \bibinfo{year}{1990}\natexlab{}.
\newblock \showarticletitle{Readings in Speech Recognition}.
\newblock \bibinfo{publisher}{Morgan Kaufmann Publishers Inc.},
  \bibinfo{address}{San Francisco, CA, USA}, Chapter A Tutorial on Hidden
  Markov Models and Selected Applications in Speech Recognition,
  \bibinfo{pages}{267--296}.
\newblock
\showISBNx{1-55860-124-4}
\urldef\tempurl%
\url{http://dl.acm.org/citation.cfm?id=108235.108253}
\showURL{%
\tempurl}


\bibitem[\protect\citeauthoryear{Radlinski and Joachims}{Radlinski and
  Joachims}{2005}]%
        {radlinski2005query}
\bibfield{author}{\bibinfo{person}{Filip Radlinski} {and}
  \bibinfo{person}{Thorsten Joachims}.} \bibinfo{year}{2005}\natexlab{}.
\newblock \showarticletitle{Query chains: learning to rank from implicit
  feedback}. In \bibinfo{booktitle}{\emph{Proceedings of the eleventh ACM
  SIGKDD international conference on Knowledge discovery in data mining}}. ACM,
  \bibinfo{pages}{239--248}.
\newblock


\bibitem[\protect\citeauthoryear{Rahman, Barson, Paul, Kayani, Lois, Quezada,
  Parnin, Stolee, and Ray}{Rahman et~al\mbox{.}}{2018}]%
        {Rahman:2018:EDU:3196398.3196425}
\bibfield{author}{\bibinfo{person}{Md~Masudur Rahman}, \bibinfo{person}{Jed
  Barson}, \bibinfo{person}{Sydney Paul}, \bibinfo{person}{Joshua Kayani},
  \bibinfo{person}{Federico~Andr{\'e}s Lois},
  \bibinfo{person}{Sebasti\'{a}n~Fernandez Quezada},
  \bibinfo{person}{Christopher Parnin}, \bibinfo{person}{Kathryn~T. Stolee},
  {and} \bibinfo{person}{Baishakhi Ray}.} \bibinfo{year}{2018}\natexlab{}.
\newblock \showarticletitle{Evaluating How Developers Use General-purpose
  Web-search for Code Retrieval}. In \bibinfo{booktitle}{\emph{Proceedings of
  the 15th International Conference on Mining Software Repositories}}
  (Gothenburg, Sweden) \emph{(\bibinfo{series}{MSR '18})}.
  \bibinfo{publisher}{ACM}, \bibinfo{address}{New York, NY, USA},
  \bibinfo{pages}{465--475}.
\newblock
\showISBNx{978-1-4503-5716-6}
\urldef\tempurl%
\url{https://doi.org/10.1145/3196398.3196425}
\showDOI{\tempurl}


\bibitem[\protect\citeauthoryear{Rao, Bansal, Mukherjee, and Maddila}{Rao
  et~al\mbox{.}}{2020}]%
        {rao2020product}
\bibfield{author}{\bibinfo{person}{Nikitha Rao}, \bibinfo{person}{Chetan
  Bansal}, \bibinfo{person}{Subhabrata Mukherjee}, {and}
  \bibinfo{person}{Chandra Maddila}.} \bibinfo{year}{2020}\natexlab{}.
\newblock \bibinfo{title}{Product Insights: Analyzing Product Intents in Web
  Search}.
\newblock
\newblock
\showeprint[arxiv]{2005.08591}~[cs.IR]


\bibitem[\protect\citeauthoryear{Rastogi, Nagappan, Gousios, and van~der
  Hoek}{Rastogi et~al\mbox{.}}{2018}]%
        {Rastogi2018}
\bibfield{author}{\bibinfo{person}{Ayushi Rastogi}, \bibinfo{person}{Nachiappan
  Nagappan}, \bibinfo{person}{Georgios Gousios}, {and}
  \bibinfo{person}{Andr\'{e} van~der Hoek}.} \bibinfo{year}{2018}\natexlab{}.
\newblock \showarticletitle{Relationship between Geographical Location and
  Evaluation of Developer Contributions in Github}. In
  \bibinfo{booktitle}{\emph{Proceedings of the 12th ACM/IEEE International
  Symposium on Empirical Software Engineering and Measurement}} (Oulu, Finland)
  \emph{(\bibinfo{series}{ESEM '18})}. \bibinfo{publisher}{Association for
  Computing Machinery}, \bibinfo{address}{New York, NY, USA}, Article
  \bibinfo{articleno}{22}, \bibinfo{numpages}{8}~pages.
\newblock
\showISBNx{9781450358231}
\urldef\tempurl%
\url{https://doi.org/10.1145/3239235.3240504}
\showDOI{\tempurl}


\bibitem[\protect\citeauthoryear{Roth and Yih}{Roth and Yih}{2005}]%
        {Roth2005}
\bibfield{author}{\bibinfo{person}{Dan Roth} {and} \bibinfo{person}{Wen-tau
  Yih}.} \bibinfo{year}{2005}\natexlab{}.
\newblock \showarticletitle{Integer Linear Programming Inference for
  Conditional Random Fields}. In \bibinfo{booktitle}{\emph{Proceedings of the
  22Nd International Conference on Machine Learning}} (Bonn, Germany)
  \emph{(\bibinfo{series}{ICML '05})}. \bibinfo{publisher}{ACM},
  \bibinfo{address}{New York, NY, USA}, \bibinfo{pages}{736--743}.
\newblock
\showISBNx{1-59593-180-5}
\urldef\tempurl%
\url{https://doi.org/10.1145/1102351.1102444}
\showDOI{\tempurl}


\bibitem[\protect\citeauthoryear{Sadowski, Stolee, and Elbaum}{Sadowski
  et~al\mbox{.}}{2015}]%
        {Sadowski2015}
\bibfield{author}{\bibinfo{person}{Caitlin Sadowski},
  \bibinfo{person}{Kathryn~T. Stolee}, {and} \bibinfo{person}{Sebastian
  Elbaum}.} \bibinfo{year}{2015}\natexlab{}.
\newblock \showarticletitle{How Developers Search for Code: A Case Study}. In
  \bibinfo{booktitle}{\emph{Proceedings of the 2015 10th Joint Meeting on
  Foundations of Software Engineering}} (Bergamo, Italy)
  \emph{(\bibinfo{series}{ESEC/FSE 2015})}. \bibinfo{publisher}{ACM},
  \bibinfo{address}{New York, NY, USA}, \bibinfo{pages}{191--201}.
\newblock
\showISBNx{978-1-4503-3675-8}
\urldef\tempurl%
\url{https://doi.org/10.1145/2786805.2786855}
\showDOI{\tempurl}


\bibitem[\protect\citeauthoryear{Sena, Coelho, Kulesza, and Bonif\'{a}cio}{Sena
  et~al\mbox{.}}{2016}]%
        {SenaMSR16}
\bibfield{author}{\bibinfo{person}{Dem\'{o}stenes Sena},
  \bibinfo{person}{Roberta Coelho}, \bibinfo{person}{Uir\'{a} Kulesza}, {and}
  \bibinfo{person}{Rodrigo Bonif\'{a}cio}.} \bibinfo{year}{2016}\natexlab{}.
\newblock \showarticletitle{Understanding the Exception Handling Strategies of
  Java Libraries: An Empirical Study}. In \bibinfo{booktitle}{\emph{Proceedings
  of the 13th International Conference on Mining Software Repositories}}
  (Austin, Texas) \emph{(\bibinfo{series}{MSR '16})}.
  \bibinfo{publisher}{Association for Computing Machinery},
  \bibinfo{address}{New York, NY, USA}, \bibinfo{pages}{212--222}.
\newblock
\showISBNx{9781450341868}
\urldef\tempurl%
\url{https://doi.org/10.1145/2901739.2901757}
\showDOI{\tempurl}


\bibitem[\protect\citeauthoryear{Shokouhi, Ozertem, and Craswell}{Shokouhi
  et~al\mbox{.}}{2016}]%
        {Shokouhi2016}
\bibfield{author}{\bibinfo{person}{Milad Shokouhi}, \bibinfo{person}{Umut
  Ozertem}, {and} \bibinfo{person}{Nick Craswell}.}
  \bibinfo{year}{2016}\natexlab{}.
\newblock \showarticletitle{Did You Say U2 or YouTube?: Inferring Implicit
  Transcripts from Voice Search Logs} \emph{(\bibinfo{series}{WWW '16})}.
  \bibinfo{pages}{1215--1224}.
\newblock
\showISBNx{978-1-4503-4143-1}
\urldef\tempurl%
\url{https://doi.org/10.1145/2872427.2882994}
\showDOI{\tempurl}


\bibitem[\protect\citeauthoryear{Sim, Umarji, Ratanotayanon, and Lopes}{Sim
  et~al\mbox{.}}{2011}]%
        {Sim2011}
\bibfield{author}{\bibinfo{person}{Susan~Elliott Sim}, \bibinfo{person}{Medha
  Umarji}, \bibinfo{person}{Sukanya Ratanotayanon}, {and}
  \bibinfo{person}{Cristina~V. Lopes}.} \bibinfo{year}{2011}\natexlab{}.
\newblock \showarticletitle{How Well Do Search Engines Support Code Retrieval
  on the Web?}
\newblock \bibinfo{journal}{\emph{ACM Trans. Softw. Eng. Methodol.}}
  \bibinfo{volume}{21}, \bibinfo{number}{1}, Article \bibinfo{articleno}{4}
  (\bibinfo{date}{Dec.} \bibinfo{year}{2011}), \bibinfo{numpages}{25}~pages.
\newblock
\showISSN{1049-331X}
\urldef\tempurl%
\url{https://doi.org/10.1145/2063239.2063243}
\showDOI{\tempurl}


\bibitem[\protect\citeauthoryear{Sindhgatta}{Sindhgatta}{2006}]%
        {Sindhgatta2006}
\bibfield{author}{\bibinfo{person}{Renuka Sindhgatta}.}
  \bibinfo{year}{2006}\natexlab{}.
\newblock \showarticletitle{Using an Information Retrieval System to Retrieve
  Source Code Samples}. In \bibinfo{booktitle}{\emph{Proceedings of the 28th
  International Conference on Software Engineering}} (Shanghai, China)
  \emph{(\bibinfo{series}{ICSE '06})}. \bibinfo{publisher}{ACM},
  \bibinfo{address}{New York, NY, USA}, \bibinfo{pages}{905--908}.
\newblock
\showISBNx{1-59593-375-1}
\urldef\tempurl%
\url{https://doi.org/10.1145/1134285.1134448}
\showDOI{\tempurl}


\bibitem[\protect\citeauthoryear{Stolee, Elbaum, and Dobos}{Stolee
  et~al\mbox{.}}{2014}]%
        {Stolee:2014:SSS:2628068.2581377}
\bibfield{author}{\bibinfo{person}{Kathryn~T. Stolee},
  \bibinfo{person}{Sebastian Elbaum}, {and} \bibinfo{person}{Daniel Dobos}.}
  \bibinfo{year}{2014}\natexlab{}.
\newblock \showarticletitle{Solving the Search for Source Code}.
\newblock \bibinfo{journal}{\emph{ACM Trans. Softw. Eng. Methodol.}}
  \bibinfo{volume}{23}, \bibinfo{number}{3}, Article \bibinfo{articleno}{26}
  (\bibinfo{date}{June} \bibinfo{year}{2014}), \bibinfo{numpages}{45}~pages.
\newblock
\showISSN{1049-331X}
\urldef\tempurl%
\url{https://doi.org/10.1145/2581377}
\showDOI{\tempurl}


\bibitem[\protect\citeauthoryear{Stylos and Myers}{Stylos and Myers}{2006}]%
        {stylos2006mica}
\bibfield{author}{\bibinfo{person}{Jeffrey Stylos} {and}
  \bibinfo{person}{Brad~A Myers}.} \bibinfo{year}{2006}\natexlab{}.
\newblock \showarticletitle{Mica: A web-search tool for finding api components
  and examples}. In \bibinfo{booktitle}{\emph{Visual Languages and
  Human-Centric Computing (VL/HCC'06)}}. IEEE, \bibinfo{pages}{195--202}.
\newblock


\bibitem[\protect\citeauthoryear{{Vasilescu}, {Filkov}, and
  {Serebrenik}}{{Vasilescu} et~al\mbox{.}}{2013}]%
        {B_Vasilescu2013}
\bibfield{author}{\bibinfo{person}{B. {Vasilescu}}, \bibinfo{person}{V.
  {Filkov}}, {and} \bibinfo{person}{A. {Serebrenik}}.}
  \bibinfo{year}{2013}\natexlab{}.
\newblock \showarticletitle{StackOverflow and GitHub: Associations between
  Software Development and Crowdsourced Knowledge}. In
  \bibinfo{booktitle}{\emph{2013 International Conference on Social
  Computing}}. \bibinfo{pages}{188--195}.
\newblock


\bibitem[\protect\citeauthoryear{Viera and Garrett}{Viera and Garrett}{2005}]%
        {kappa}
\bibfield{author}{\bibinfo{person}{Anthony~J. Viera} {and}
  \bibinfo{person}{Joanne~M. Garrett}.} \bibinfo{year}{2005}\natexlab{}.
\newblock \showarticletitle{Understanding interobserver agreement: the kappa
  statistic.}
\newblock \bibinfo{journal}{\emph{Family Medicine}}  \bibinfo{volume}{37.5}
  (\bibinfo{year}{2005}), \bibinfo{pages}{360--363}.
\newblock


\bibitem[\protect\citeauthoryear{Wang}{Wang}{2009}]%
        {Wang09}
\bibfield{author}{\bibinfo{person}{Yefeng Wang}.}
  \bibinfo{year}{2009}\natexlab{}.
\newblock \showarticletitle{Annotating and Recognising Named Entities in
  Clinical Notes}. In \bibinfo{booktitle}{\emph{{ACL} 2009, Proceedings of the
  47th Annual Meeting of the Association for Computational Linguistics and the
  4th International Joint Conference on Natural Language Processing of the
  AFNLP, 2-7 August 2009, Singapore, Student Research Workshop}}.
  \bibinfo{pages}{18--26}.
\newblock


\bibitem[\protect\citeauthoryear{Wei, Chen, Xu, He, and Gui}{Wei
  et~al\mbox{.}}{2016}]%
        {wei2016disease}
\bibfield{author}{\bibinfo{person}{Qikang Wei}, \bibinfo{person}{Tao Chen},
  \bibinfo{person}{Ruifeng Xu}, \bibinfo{person}{Yulan He}, {and}
  \bibinfo{person}{Lin Gui}.} \bibinfo{year}{2016}\natexlab{}.
\newblock \showarticletitle{Disease named entity recognition by combining
  conditional random fields and bidirectional recurrent neural networks}.
\newblock \bibinfo{journal}{\emph{Database}}  \bibinfo{volume}{2016}
  (\bibinfo{year}{2016}).
\newblock


\bibitem[\protect\citeauthoryear{Xia, Bao, Lo, Kochhar, Hassan, and Xing}{Xia
  et~al\mbox{.}}{2017}]%
        {Xia:2017}
\bibfield{author}{\bibinfo{person}{Xin Xia}, \bibinfo{person}{Lingfeng Bao},
  \bibinfo{person}{David Lo}, \bibinfo{person}{Pavneet~Singh Kochhar},
  \bibinfo{person}{Ahmed~E. Hassan}, {and} \bibinfo{person}{Zhenchang Xing}.}
  \bibinfo{year}{2017}\natexlab{}.
\newblock \showarticletitle{What do developers search for on the web?}
\newblock \bibinfo{journal}{\emph{Empirical Software Engineering}}
  \bibinfo{volume}{22} (\bibinfo{date}{04} \bibinfo{year}{2017}).
\newblock
\urldef\tempurl%
\url{https://doi.org/10.1007/s10664-017-9514-4}
\showDOI{\tempurl}


\bibitem[\protect\citeauthoryear{Zhang, Jain, Khandelwal, Kaushik, Ge, and
  Hu}{Zhang et~al\mbox{.}}{2016}]%
        {Zhang2016}
\bibfield{author}{\bibinfo{person}{Hongyu Zhang}, \bibinfo{person}{Anuj Jain},
  \bibinfo{person}{Gaurav Khandelwal}, \bibinfo{person}{Chandrashekhar
  Kaushik}, \bibinfo{person}{Scott Ge}, {and} \bibinfo{person}{Wenxiang Hu}.}
  \bibinfo{year}{2016}\natexlab{}.
\newblock \showarticletitle{Bing Developer Assistant: Improving Developer
  Productivity by Recommending Sample Code}. In
  \bibinfo{booktitle}{\emph{Proceedings of the 2016 24th ACM SIGSOFT
  International Symposium on Foundations of Software Engineering}} (Seattle,
  WA, USA) \emph{(\bibinfo{series}{FSE 2016})}. \bibinfo{publisher}{Association
  for Computing Machinery}, \bibinfo{address}{New York, NY, USA},
  \bibinfo{pages}{956--961}.
\newblock
\showISBNx{9781450342186}
\urldef\tempurl%
\url{https://doi.org/10.1145/2950290.2983955}
\showDOI{\tempurl}


\end{thebibliography}

\end{document}